\documentclass[useAMS,usenatbib,fleqn]{mn2e}

\usepackage{times}

\usepackage{deluxetable}
\usepackage{graphicx}
\usepackage{epsfig}
\usepackage{amssymb,amsmath}
\usepackage{aas_macros}
\usepackage{fixltx2e} 
\title[Tilted disc models of Sgr A*]{Tilted black hole accretion disc
  models of Sagittarius A*:\\ time-variable millimetre to near-infrared emission}
\author[Dexter \& Fragile]{Jason Dexter$^{1}$\thanks{E-mail:
    jdexter@berkeley.edu} and P. Chris Fragile$^{2}$\\
$^{1}$Theoretical Astrophysics Center and Department of Astronomy,
University of California, Berkeley, CA 94720-3411, USA\\
$^{2}$Department of Physics \& Astronomy, College of Charleston, Charleston, SC 29424, USA}
\begin{document}

\pagerange{\pageref{firstpage}--\pageref{lastpage}} \pubyear{2012}
\maketitle

\label{firstpage}

\begin{abstract}
High-resolution, multi-wavelength, and time-domain observations of the
Galactic centre black hole 
candidate, Sgr A*, allow for a direct test of contemporary accretion
theory. To date, all models have assumed 
alignment between the accretion disc and black hole angular momentum
axes, but this is unjustified for geometrically thick
accretion flows like that onto Sgr A*. Instead, we calculate images and spectra
from a set of numerical simulations of accretion flows
misaligned (``tilted'') by $15^\circ$ from the black hole spin axis and compare
them with millimetre (mm) to near-infrared (NIR) observations. Non-axisymmetric
standing shocks from eccentric fluid orbits dominate the
emission, leading to a wide range of possible image
morphologies. These effects invalidate previous parameter estimates from
model fitting, including estimates of the dimensionless black hole
spin, except possibly for very low values of spin or tilt (upper
limits of $a < 0.3$ or $\beta < 15^\circ$). At $1.3$mm, the black hole
images still have crescent morphologies, and the black hole shadow may
still be accessible to future very long baseline interferometry
(mm-VLBI) observations. Shock heating leads to multiple populations of
electrons, some at high energies ($T_e > 10^{12} \rm K$). These
electrons can naturally produce the observed NIR flux, spectral index,
and rapid variability (``flaring''). This NIR emission is uncorrelated
with that in the mm, which also agrees with observations. 

These are the first models to self-consistently explain the time-variable mm to
NIR emission of Sgr A*. Predictions of the model include 
significant structural changes observable with mm-VLBI on both the
dynamical (hour) and Lense-Thirring precession (day-year) timescales; 
and $\simeq 30-50 \mu$as changes in centroid position from extreme
gravitational lensing events during NIR flares, detectable with the
future VLT instrument GRAVITY. We further predict that multi-wavelength 
monitoring should find no significant correlations between mm and
NIR/X-ray light curves. The weak correlations reported to date are
shown to be consistent with our model, where they are artifacts of the short
light curve durations. If the observed NIR emission is caused by
shock heating in a tilted accretion disc, then the Galactic centre black hole
has a positive, non-zero spin parameter ($a > 0$). 
\end{abstract}


\section{Introduction}

Due to its proximity, Sgr A* is the most intensively studied
supermassive black hole candidate and the largest in angular
size. Recent very long baseline interferometry observations at 1.3mm
\citep[mm-VLBI][]{doeleman2008,fishetal2011} have detected source
structure on event-horizon scales (tens of microarcseconds,
$\mu$as). Future measurements with this Event Horizon Telescope may
detect the black hole shadow, providing the first direct evidence for
an event horizon \citep{bardeen1973,falcke,dexteretal2010}.

The spectral energy distribution of Sgr A* rises from the radio to a
peak in the submillimetre (mm ``bump''). The radio emission is
synchrotron radiation from non-thermal electrons far from the black hole, either
in the accretion flow \citep{yuanquataert2003} or in a mildly
relativistic jet \citep{falckemarkoff2000}. The mm bump is
well described by synchrotron radiation from hot ($\sim 10^{10-11}
\rm K$), mostly thermal electrons in the immediate vicinity of the
black hole \citep[$r \simeq 5-10 \rm M$\footnote{We frequently use
  units with $G=c=1$; in these units for Sgr A* $1 \rm M$ is
  approximately $6\times10^{11} \rm cm$ and $20 \rm
  s$.}][]{yuanquataert2003,moscibrodzka2009,dexteretal2010}.

In addition to $30-50 $ per cent variability in the mm
\citep{zhao2003,yusefzadehetal2006,eckartetal2006,marrone2008}, Sgr A*
is observed to ``flare'' rapidly in the NIR
\citep{genzel2003,ghez2004} and the X-ray \citep{baganoff2001} with
order of magnitude amplitudes. The NIR/X-ray flares occur
simultaneously \citep{eckartetal2004}, and weak
correlations in simultaneous light curves have suggested possible lags
between NIR/X-ray and mm variability \citep{yusefzadehetal2008}. The
NIR (X-ray) emission is either synchrotron radiation from energetic
electrons, or inverse Compton scattering of mm (NIR) seed photons
\citep{markoffetal2001}. 

No theoretical model of Sgr A* self-consistently accounts for the observed
time-variable multi-wavelength emission. Radiatively ineffecient
accretion flow \citep[RIAF][]{yuanquataert2003} and jet
\citep{falckemarkoff2000} models can describe the multi-wavelength
``quiescent'' spectrum. RIAF models also provide excellent fits to
existing mm-VLBI data \citep{broderick2009,brodericketal2011}. 

Models for the NIR flares include coherently orbiting inhomogeneities
\citep[``hotspots''][]{broderickloeb2006} in the inner radii of the
accretion flow. Recently, correlated multiwavelength flares have
instead favored an adiabatically expanding blob model
\citep{yusefzadeh2009,eckartetal2012}, which can explain the reported
time lags between the NIR/X-ray and mm flares. The blobs could be
formed from non-thermal events such as magnetic reconnection in the
accretion flow \citep{doddsedenetal2010}. 

All of these models are non-relativistic. They neglect the magnetic
fields responsible for angular momentum transport and accretion via
the magnetorotational instability
\citep[MRI,][]{mri}. Magnetohydrodynamic (MHD) simulations can provide
a more physical description of the accretion flow. They have been used
to model the synchrotron emission from Sgr A*, either in three spatial
dimensions with a pseudo-Newtonian potential
\citep{ohsuga2005,goldston2005,huang2009,chan2009}, or in full general
relativity (GRMHD) in two
\citep{noble2007,moscibrodzka2009} or three
\citep{dexter2009,dexteretal2010,shcherbakovetal2012,dolenceetal2012}
dimensions. Non-relativistic 
simulations are especially inappropriate for modeling the mm/NIR/X-ray  
emission, which originates in the innermost portion of the accretion
flow where relativistic effects are strongest. Axisymmetric
simulations cannot sustain the MRI, and cannot accurately model
variability. 

Existing models from 3D simulations provide an excellent
description of mm observations including polarization
\citep{shcherbakovetal2012}, variability and mm-VLBI
\citep{dexter2009,dexteretal2010}. These models also produce large
amplitude NIR flares \citep{dolenceetal2012}, but with much lower
fluxes and steeper spectral indices than observed. Although current
global GRMHD simulations are factors of a few in resolution from
convergence \citep{hawleyetal2011}, the resulting spectra of Sgr A*
are fairly insensitive to resolution \citep{shiokawaetal2012}. 

There are two major limitations to current models. First, the
electron distribution function is highly uncertain. Most theoretical
models adopt a thermal distribution for the mm bump, and some add a
non-thermal power law tail. Models from simulations use a thermal
distribution with a constant electron-ion temperature ratio,
$T_i/T_e$, allowing the electron temperature to be calculated from the
output density and pressure
\citep{goldston2005,moscibrodzka2009}. Simple prescriptions for
allowing this ratio to vary spatially lead to only minor changes in
the images/spectra. This parameter can be calibrated either using
physically motivated sub-grid models for electron heating
\citep{sharma2007e} or through direct particle-in-cell calculations of
MRI turbulence \citep{riquelmeetal2012}. The second major limitation to all existing
theoretical models of Sgr A*, and the subject of this article, is the
assumption that the accretion flow angular momentum axis is aligned
with the black hole spin axis.  

Geometrically thick accretion flows like that onto Sgr A* do not align with a central black
hole as geometrically thin discs do
\citep{bardeenpetterson1975}. Sgr A* is currently fed by the winds of
young, massive stars \citep{quataert2004}, whose net angular momentum
axis is not expected to be aligned with the spin axis of Sgr A*. 
The only way for alignment to occur is then 
for the black hole to accrete an amount of material with constant
angular momentum orientation comparable to
its mass. For this to occur in a Hubble time requires an accretion
rate $\sim 10^{-4} \dot{M} \rm yr^{-1}$, an order of magnitude larger
than the estimated accretion rate at the Bondi radius
\citep{yuanquataert2003} and several orders of magnitude larger than
that onto Sgr A* \citep[$10^{-7}-10^{-9} M_\odot \rm
yr^{-1}$][]{aitken2000,bower03,marrone07}. Therefore, it is highly
likely that the accretion flow is misaligned (``tilted'').

The expected disc misalignment has extraordinary consequences for the
dynamics and observable properties of a black hole accretion
flow. Lense-Thirring precession leads to tilts and warps that are
oscillatory functions of radius \citep{ivanov1997}, and gravitational
torques cause the entire disc to precess essentially as a solid body
\citep{fragile2007}. Eccentric orbits in the disc
converge, leading to non-axisymmetric standing shocks
\citep{fragile2008}. Excess angular momentum transport from the
shocks truncates the accretion flow outside of the marginally
stable orbit \citep{fragiletilt2009,dexterfragile2011}. The asymmetry
of the accretion flow leads to images and spectra that depend strongly
on the observer azimuthal viewing angle \citep{dexterfragile2011}. 

In this work we calculate time-dependent images and spectra 
from the only published GRMHD simulations of tilted black hole
accretion discs \citep{fragile2007,fragileetal2009,fragiletilt2009}, and compare them
with mm to NIR observations of Sgr A*. We review the physics of tilted accretion
discs in \S \ref{sec:tilted-black-hole} and describe the numerical simulations 
in \S \ref{sec:sims}. The methods used to construct radiative models from
the simulations are outlined in \S
\ref{modeling}. These radiative models are then fit to existing
mm-VLBI and spectral observations in \S \ref{fitting}, and the mm
images and mm to NIR spectral and variable properties of the
best-fitting models are analyzed in \S \ref{sec:model-properties}. The
implications and limitations of this study are discussed in \S
\ref{sec:discussion}, and the major results are summarized in \S \ref{sec:summary}.

\begin{figure*}
\includegraphics[scale=.8]{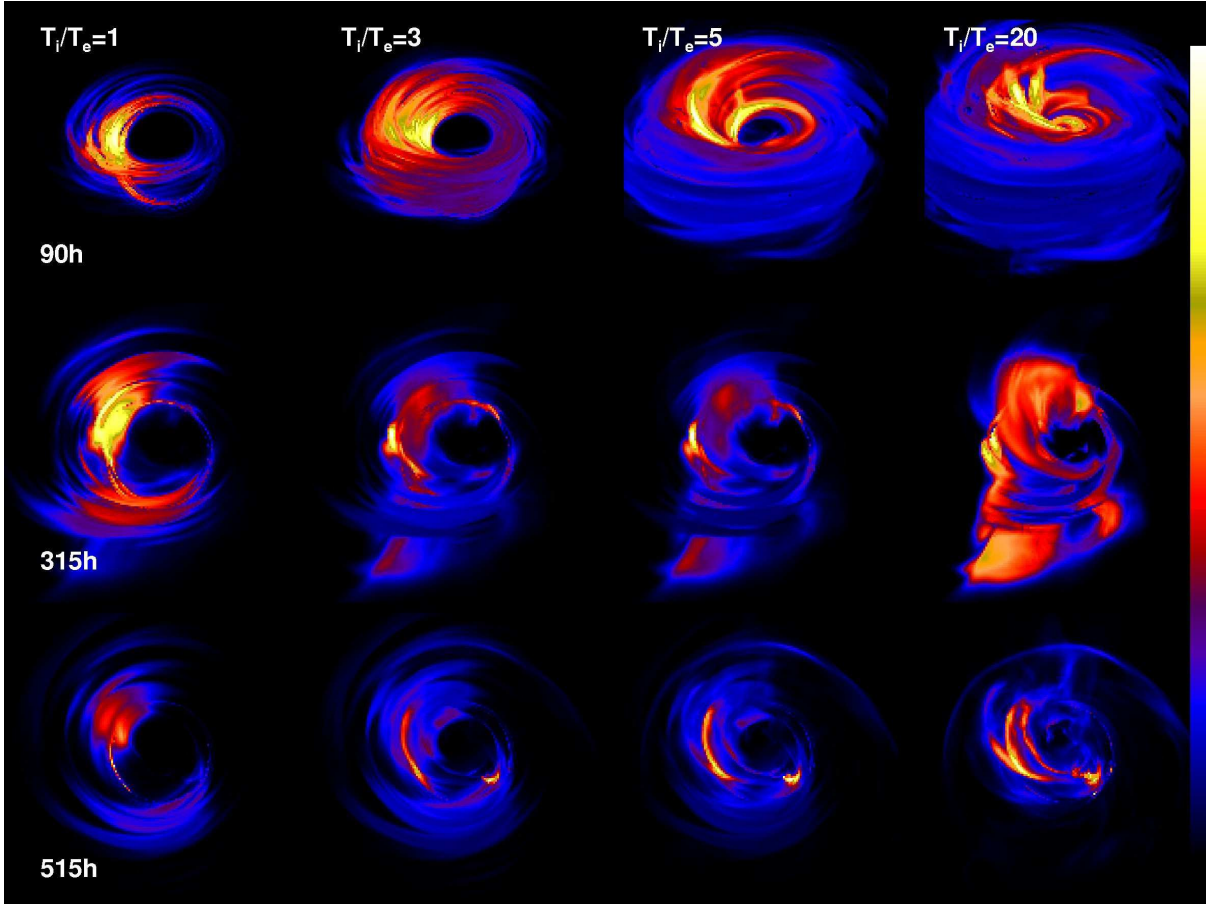}
\caption{\label{titeimg}Images as a function of $T_i/T_e$ for the 90h,
315h, and 515h simulations. The colours are scaled linearly with a
dynamic range of 60, and the images are $160\times160 \mu$as in all
cases. Images from untilted simulations such as
90h become optically thick and emit from a growing photosphere with
increasing $T_i/T_e$. Images from
tilted simulations do not become uniformly optically thick, and can
even become smaller at large $T_i/T_e$, as in the case of 515h, due to
the presence of multiple electron populations with varying electron
temperatures at distinct spatial locations in the accretion flow.}
\end{figure*}

\begin{figure*}
\begin{center}
\begin{tabular}{ll}
\includegraphics[scale=.7]{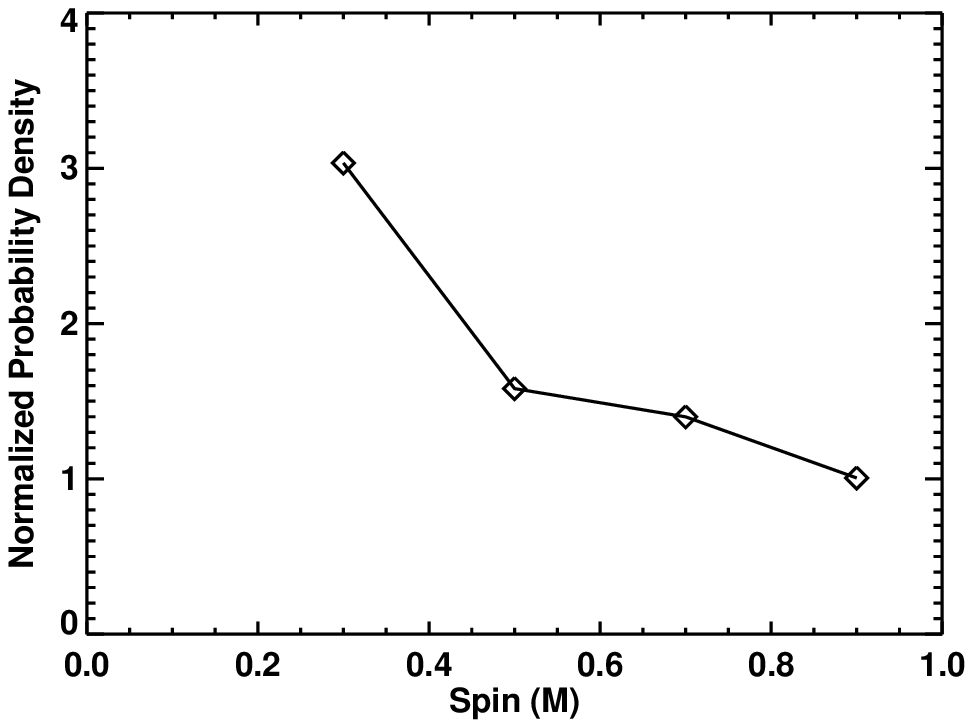}&\includegraphics[scale=.7]{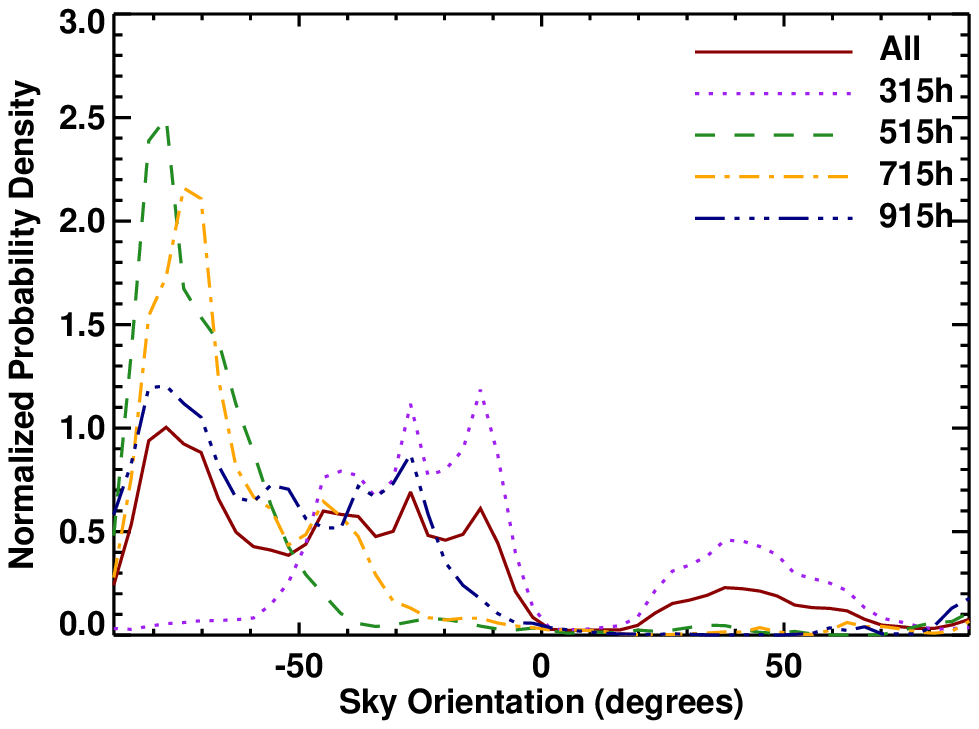}\\
\includegraphics[scale=.7]{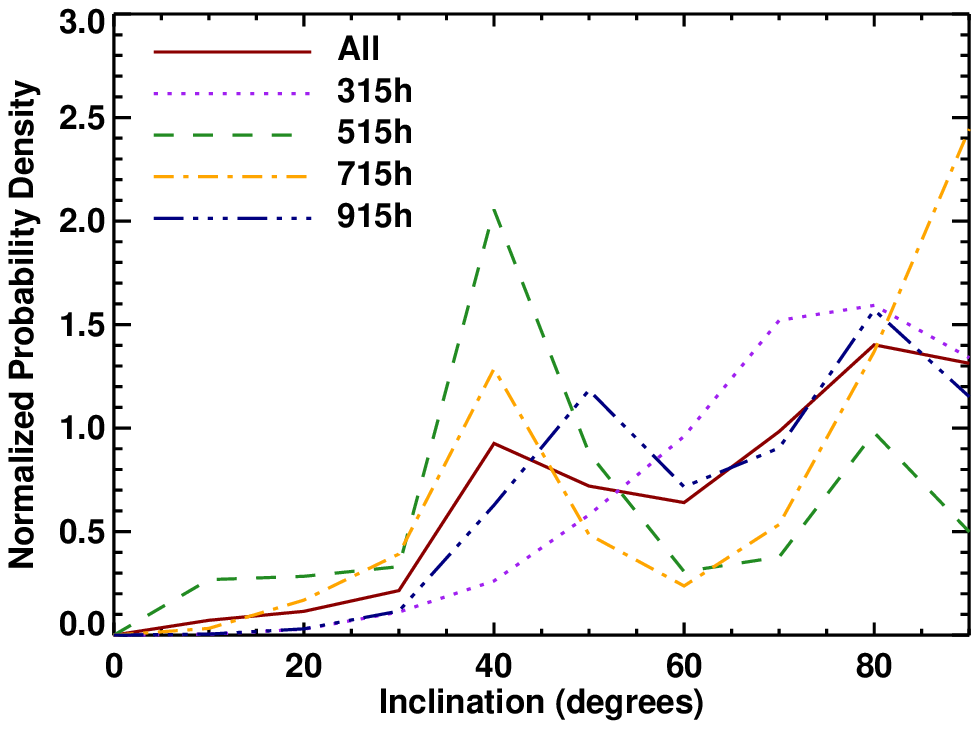}&\includegraphics[scale=.7]{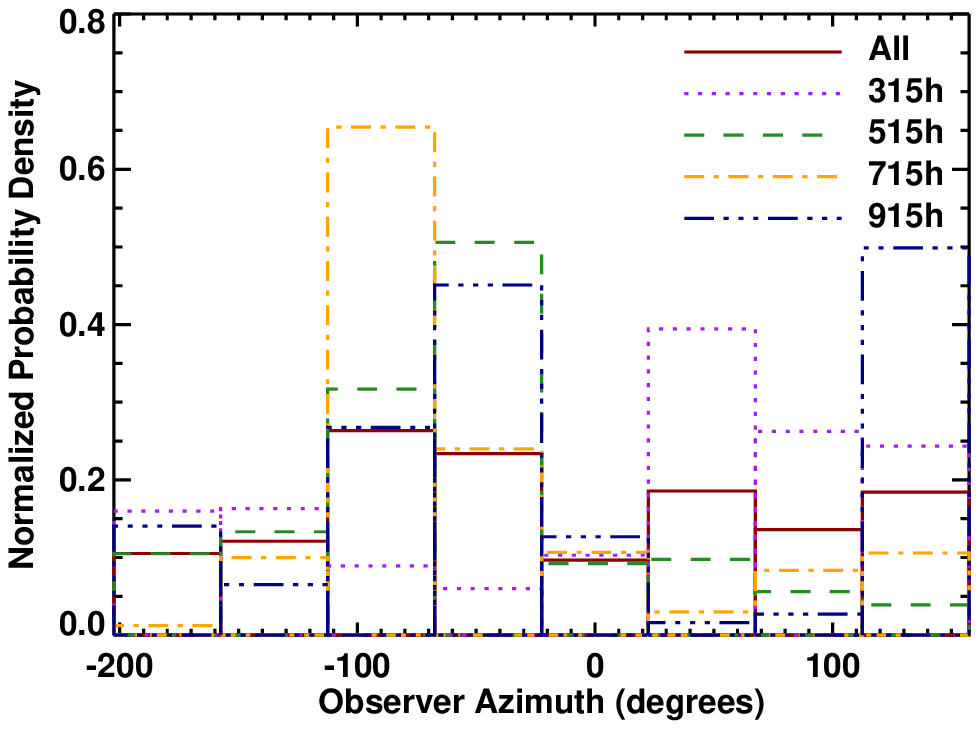}\\
\end{tabular}
\caption{\label{probdiststilti}Normalized probability distributions as
  functions of black hole spin (top left), observer inclination (top
  right), sky orientation (bottom left), and observer azimuth (bottom
  right) for each simulation separately and combined (solid
  lines). Unlike in aligned models \citep{broderick2009,brodericketal2011,dexteretal2010} the parameters of the viewing
  geometry are essentially unconstrained.}
\end{center}
\end{figure*}

\section{Tilted Black Hole Accretion Discs}
\label{sec:tilted-black-hole}

Whenever a black hole accretion disc is tilted, it experiences
differential Lense-Thirring precession owing to the frame dragging of
the rotating black hole. In a disc, Lense-Thirring
precession is able to build up over many orbital periods, so it can
affect the structure well beyond radii normally associated with
relativistic effects.

The effect of Lense-Thirring precession in a disc is to cause it to
warp, and this warping can propagate through the disc in either a
diffusive or wave-like manner. In the diffusive case, the warping is
limited by ``viscous'' responses within the disc, and the
Lense-Thirring precession dominates out to a unique transition radius
$r_\mathrm{BP}$ \citep{bardeenpetterson1975,kumarpringle1985}, inside
of which the disc is expected to be flat and aligned with the
black-hole midplane, and outside of which it is also expected to be
flat but in a plane determined by the angular momentum of the gas
reservoir. This ``Bardeen-Petterson'' configuration is expected for Keplerian discs whenever the
dimensionless stress parameter $\alpha$ is larger than the ratio
$H/r$, where $H$ is the disc semi-thickness. Given that  $\alpha$ is
usually considered to be significantly less than one, this requires
geometrically thin discs.

In thicker discs, such as would be expected in low-luminosity sources
such as Sgr A*, warps propagate in a wave-like manner, and the disc
does not go into the Bardeen-Petterson configuration.  Instead, the
tilt of the disc can be an oscillatory function of radius, with the
amplitude of the oscillations growing as one approaches the black hole
\citep{ivanov1997}.  These oscillations are stationary, however, so
they cannot act to absorb the torque of the black hole.  For thick
discs, this torque instead causes the disc to precess as if it 
were a solid body. 

The warping of tilted thick discs creates unbalanced
pressure gradients that can drive significant latitude-dependent
epicyclic motion within the disc. The induced motion of the gas 
can be coherent over the entire scale of the disc, with magnitudes that
are substantial fractions of the orbital velocity.  Radial variations
in the eccentricity of the associated fluid-element trajectories can
lead to a crowding of orbit trajectories at certain locations within
the disc. This results in local density enhancements akin to
compressions. These compressions can be sufficiently strong to produce
standing shocks within the disc, particularly in the vicinity of the
black hole \citep{fragile2008}.

Tilted accretion disc models can be constructed semi-analytically
using height-integrated equations in the so-called twisted frame
\citep{ivanov1997,zhuravlevivanov2011}. However, these models lack the magnetic
fields that lead to angular momentum transport in the disc and give
rise to the observed synchrotron emission from Sgr A*. They also have
an unconstrained vertical structure, leading to additional free
parameters. Finally, they are stationary, and so cannot be used to
study variability self-consistently. Instead, we employ 3D GRMHD
simulations of misaligned black hole accretion discs. Simulations have
fewer free parameters and higher physical fidelity. Few exist,
however, preventing anything like a thorough exploration of the
parameter space in dimensionless black hole spin, disc tilt, and geometric disc
thickness. 

\begin{table}
\caption{Simulation Parameters \label{sims}}
\begin{minipage}{6cm}
\begin{small}
\begin{center}
\begin{tabular}{lccc}

        \tableline
Simulation & $a$ & $\beta$ & Reference\\
        \tableline
315h & 0.3 & $15^\circ$ & \citet{fragiletilt2009}\\
515h & 0.5 & $15^\circ$ & \citet{fragileetal2009}\\
715h & 0.7 & $15^\circ$ & \citet{fragiletilt2009}\\
90h & 0.9 & $0^\circ$ & \citet{fragile2007}\\
915h & 0.9 & $15^\circ$ & \citet{fragile2007}\\
	\tableline
\end{tabular}
\end{center}
\end{small}
\end{minipage}
\end{table}

\begin{figure*}
\begin{center}
\begin{tabular}{ll}
\includegraphics[scale=.7]{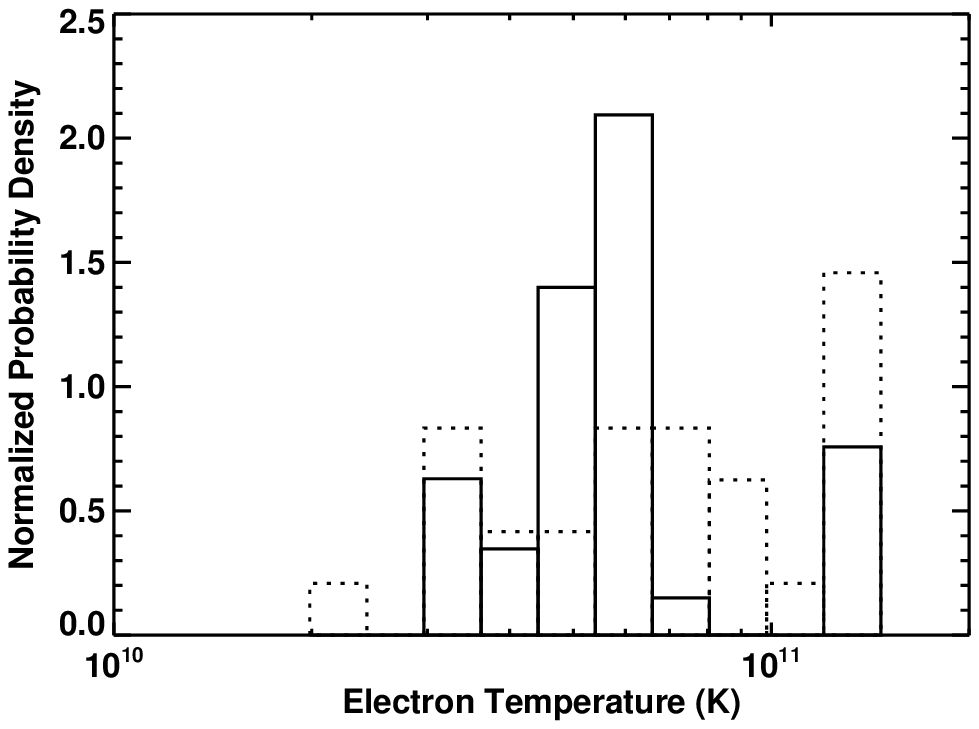}&\includegraphics[scale=.7]{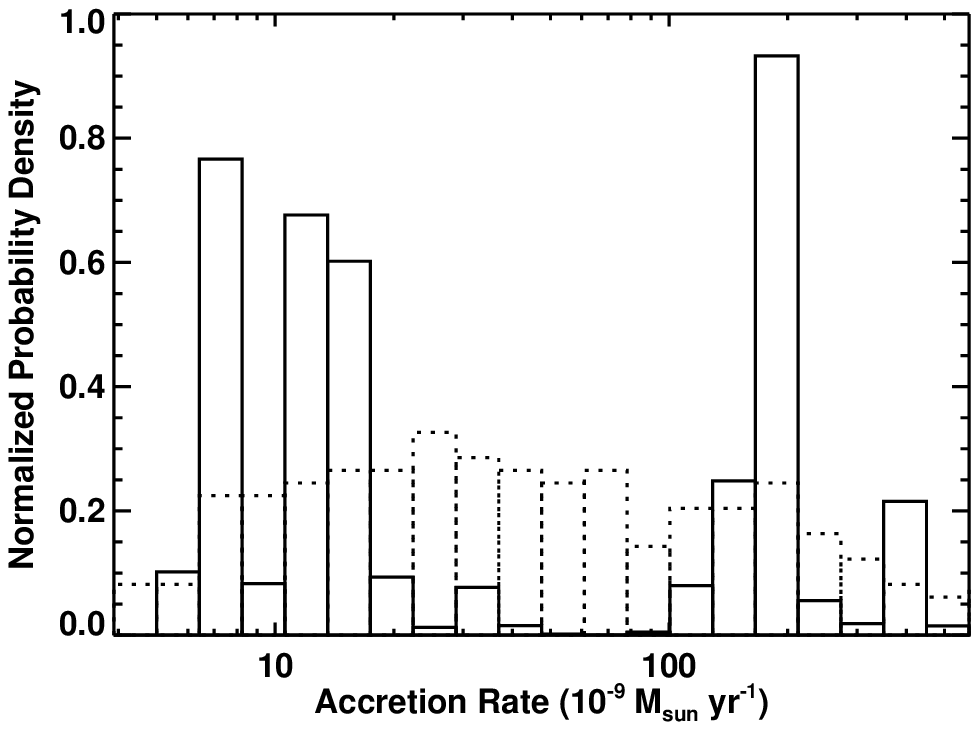}
\end{tabular}
\caption{\label{probtitemdot}Normalized probability density as a function of
  electron temperature (left) and accretion rate (right), marginalized
  over the other parameters. The electron temperature is taken to be
  the median in the region of largest emissivity for each
  model. The accretion rate is time-averaged for
  each model. In both cases, probability densities for models falling in each bin are
  added to make the histogram. The dotted histograms in both panels
  are the result of assigning equal probability to each model. The
  viable range of accretion rates is much larger than in aligned
  simulations \citep{dexter2009,dexteretal2010}, while the electron
  temperature remains well constrained to a narrow region around the
  observed brightness temperature.}
\end{center}
\end{figure*}

\section{Simulations}
\label{sec:sims}

The numerical data used in our analysis were taken from simulations
presented in \citet{fragile2007}, \citet{fragileetal2009}, and \citet{fragiletilt2009}. All of
the simulations used the Cosmos++ GRMHD code \citep{anninos2005}. 
Cosmos++ includes several schemes for solving the GRMHD equations;
in these simulations, the internal-energy, artificial viscosity formulation was used. The
magnetic fields were evolved in an advection-split form, while using
a hyperbolic divergence cleanser to maintain an approximately
divergence-free magnetic field. The GRMHD equations were evolved in
a ``tilted'' Kerr-Schild polar coordinate system
$({t},{r},{\vartheta},{\varphi})$. This coordinate system is related
to the usual (untilted) Kerr-Schild coordinates
$({t},{r},{\theta},{\phi})$ through a simple rotation about the
${y}$-axis by an angle $\beta$, as described in \citet{fragile2005}.

The simulations were carried out on a spherical polar mesh with
nested resolution layers. The base grid contained $32^3$ mesh zones
and covered the full $4\pi$ steradians. Two levels of refinement
were added on top of the base layer; each refinement level doubling
the resolution relative to the previous layer, thus
achieving peak resolutions equivalent to a $128^3$ simulation. 
In the radial direction a logarithmic coordinate of the form $\eta
\equiv 1.0 + \ln (r/r_{\rm BH})$ was used, where $r_{\rm BH}$ is the black-hole horizon radius. 
In the angular direction, in addition to
the nested grids, a concentrated latitude coordinate $x_2$ of the
form $\vartheta = x_2 + \frac{1}{2} (1 - h) \sin (2 x_2)$ was used
with $h = 0.5$ to concentrate resolution toward the midplane of
the disc. 

The simulations started from the analytic solution for an
axisymmetric torus around a rotating black hole \citep{chakrabarti1985}. The inner radius of
the torus was $r_{\rm in}=15 M$; the radius of the initial
pressure maximum was $r_{\rm centre}=25 M$; and the
power-law exponent used in defining the initial specific angular
momentum distribution was $q=1.68$. An adiabatic equation of state
was assumed, with $\Gamma=5/3$. The torus was seeded with a weak
dipole magnetic field in the form of poloidal loops along the
isobaric contours within the torus. The field was normalized such
that initially $\beta_{\rm mag} =P/P_B \ge \beta_{\rm mag,0}=10$
throughout the torus. The black
hole was inclined by an angle $\beta=0^\circ$ or $15^\circ$ relative to the
disc (and the grid).
From this starting point, the simulations were allowed to
evolve for a time equivalent to 10 orbits at the initial pressure
maximum, $r_{\rm centre}$, corresponding to hundreds of orbits at
the ISCO. Table \ref{sims} summarizes the simulation parameters.

\section{Radiative Modeling}

A radiative model is specified by three components: a dynamical model
for fluid variables as functions of space and time; a model for the
electron distribution function; and an emission model. In this work,
the GRMHD simulations described above are used as the dynamical
model. As in previous radiative models of Sgr A* from simulations
\citep{goldston2005,moscibrodzka2009,dexteretal2010}, we employ a
thermal Maxwell-Juttner distribution for the electrons, assuming a
constant ion-electron temperature ratio, $T_i/T_e$, as a crude
approximation to the electron distribution function in a
collisionless plasma. This ratio is a free parameter in the
models. Finally, we assume the emission is entirely from synchrotron
radiation, and use the unpolarized emission coefficient
from \citet{leungetal2011}.

\label{modeling}
With these assumptions, observables can be calculated from the
simulations using ray tracing. Starting from an observer's camera, rays are
traced backwards in time toward the black hole (assuming they are null
geodesics) using the public code
\textsc{geokerr} 
\citep{dexteragol2009}. In the region where rays intersect the
accretion flow, the radiative transfer equation is solved along the
geodesic using the code \textsc{grtrans} \citep{dexter2011}, which
then represents a pixel of the image. This 
procedure is repeated for many rays to produce an image, and at many
time steps of the simulation to produce time-dependent images
(movies). Light curves are computed by integrating over the individual
images. Repeating the procedure over observed wavelengths gives a
time-dependent spectrum. 

To calculate fluid properties at each point on a ray, the spacetime
coordinates of the geodesic are transformed from Boyer-Lindquist to
the tilted Kerr-Schild coordinates used in the simulations. Since the accretion flow is dynamic, light
travel time delays along the geodesic are taken into account. Data
from the sixteen nearest zone centres (eight on the simulation grid
over two time steps) were interpolated to each point on the
geodesic. The simulations are scale free, and must be scaled to cgs units in
order to calculate emission and absorption coefficients. Fixing the
black hole mass ($M_{\rm bh} \simeq 4 \times10^{6} M_\odot$ for Sgr A*) sets the
length and timescales. The total mass in the accretion flow is a free
parameter, and its choice is equivalent to fixing the time-averaged
accretion rate. To convert image intensities to flux, we adopt a
distance $D = 8 \rm kpc$ to the Galactic centre. 

For each simulation, images are produced over a grid of 
parameters at $0.4$mm and $1.3$mm: dimensionless black hole
spin, $a$ (or equivalently simulation); time-averaged accretion rate,
$\dot{M}$; ion-electron temperature ratio, $T_i/T_e$; observer
inclination, $i$, azimuth, $\phi_0$, and sky orientation (position
angle measured E of N), $\xi$.  These images \citep[including the
effects of interstellar scattering,][]{bower2006} are fit jointly to 
mm-VLBI observations \citep[][]{doeleman2008,fishetal2011} and spectral measurements 
\citep{marronephd}. This gives the probability of observing the
measured values from a particular model, which is converted to the
probability distribution as a function of the model parameters using
the method described in \citet[][]{broderick2009} and used in
\citet{dexteretal2010}. This procedure gives a joint probability 
distribution as a function of the above parameters. To estimate
individual parameters, we marginalize over the others using uniform
priors on $a$, $T_i/T_e$, $\phi_0$, and $\xi$; a logarithmic prior on
$\dot{M}$, and a $\sin{i}$ prior on the inclination angle. 

In practice, the range of $\dot{M}$ is used to bracket the range of
observed mm (230 GHz) fluxes from Sgr A*: $\approx 2-5 \rm
Jy$. Achieving convergence requires a fine enough grid so that multiple 
images have mm fluxes close to those observed by mm-VLBI ($\simeq 2.4
\rm Jy$) and spectral observations ($\simeq 3.5-4.25 \rm Jy$). In
untilted simulations, this is possible with a fairly coarse grid of 10
$\dot{M}$ values, since the image structure is nearly unchanged as the
flux varies by $30-50 $ per cent (see
\S\ref{sec:images}). Variability then acts to sample the total flux more
finely. This is not the case in tilted simulations, where the image
structure changes drastically on short timescales. In probability distributions over both observer time
and position angle ($\xi$), reasonable convergence is obtained when
using $20$ values of $\dot{M}$. 

The observer azimuth, $\phi_0$, is a
new parameter for tilted simulations, since the accretion flow is no
longer azimuthally symmetric on times much longer than the orbital
time. Instead, non-axisymmetric structure in these simulations
averages out only on the much longer precession timescale.\footnote{The
  simulations considered here have durations of $\approx t_{\rm prec}
  / 8$.} Also unlike
untilted simulations, the non-axisymmetric structure dominates images
from tilted discs \citep{dexterfragile2011}. We find that $8$ values
of $\phi_0$ are sufficient to capture the range of model images.

\begin{figure}
\includegraphics[scale=.8]{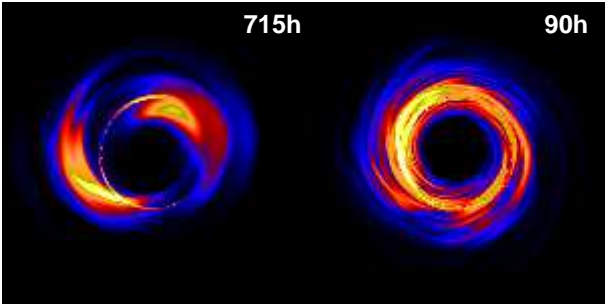}
\caption{\label{shockimgs}Nearly face-on images from the 715h (left) 
  and 90h (right) simulations scaled as in Figure \ref{titeimg}. The structure in images from untilted simulations
  such as 90h is mostly azimuthally symmetric (circularly symmetric when
  viewed face-on), while that in the tilted simulations such as 715h
  is highly non-axisymmetric due to the presence of standing shocks
  which dominate the mm emission in models of Sgr A*. In
  addition, asymmetries in untilted images travel at the orbital
  speed, whereas the standing shocks travel at the much slower
  Lense-Thirring precession speed.}
\end{figure}

\begin{figure*}
\includegraphics[scale=1]{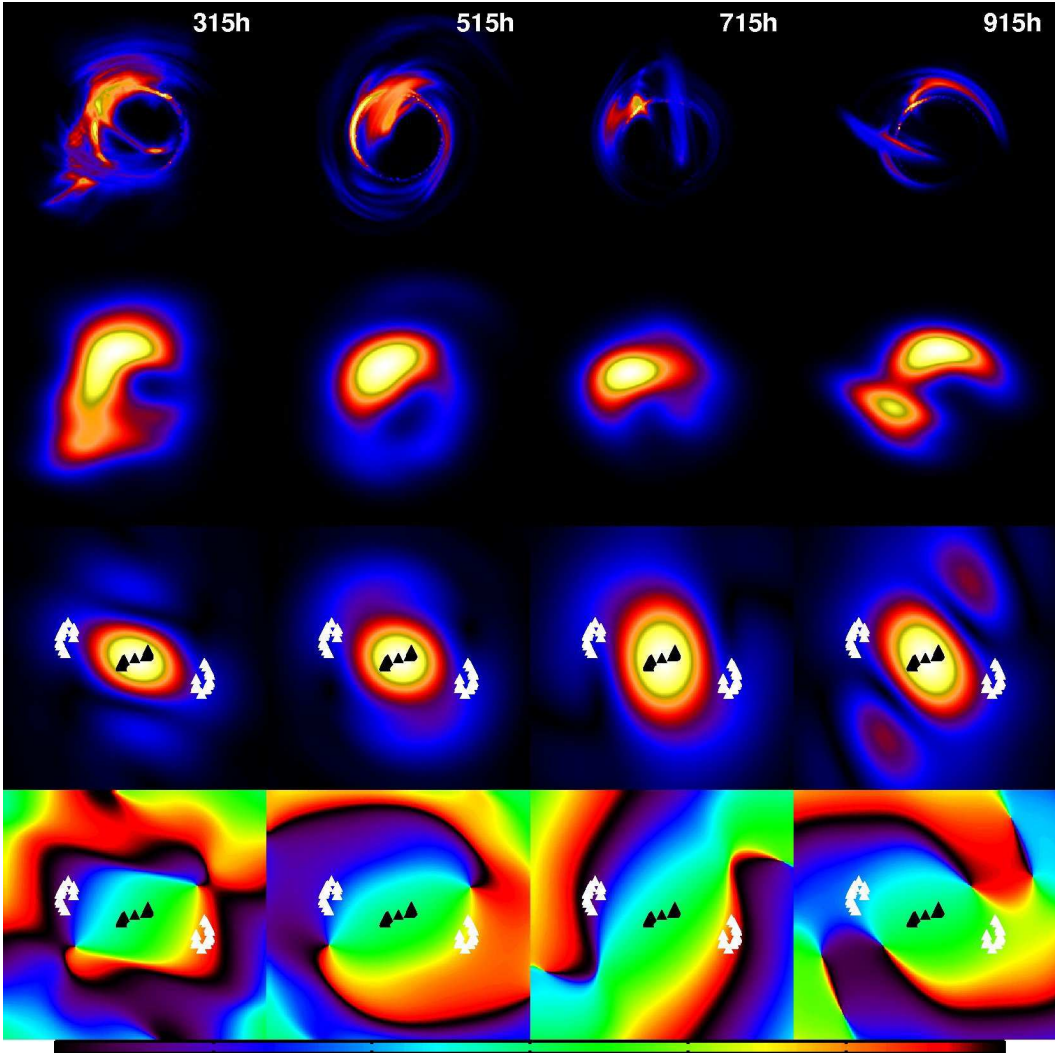}
\caption{\label{dint230}Best-fitting images from each simulation (columns)
  with (second row) and without (top row) the effects of interstellar
  scattering, as well as the corresponding visibility
  amplitudes (third row) and phases (fourth row) for $\nu_0=1.3$mm
  ($230$ GHz). Image scales are the same as in Figure
  \ref{titeimg}. The visibility amplitudes also have a dynamic range
  of 60, and both the visibility amplitude and phase images are
  $1200\times1200 M\lambda$ in size. The phase colours range from blue
  (negative) to green (zero) to red (positive). The triangles show the
  locations of existing mm-VLBI observations. Once convolved with interstellar
  scattering, the best-fitting 
  images are crescents with significant but non-dominant Doppler
  beaming. The black hole shadow appears as a local minimum in the
  visibility amplitude, and a region of rapidly varying phase.}
\end{figure*}

\begin{figure}
\includegraphics[scale=.8]{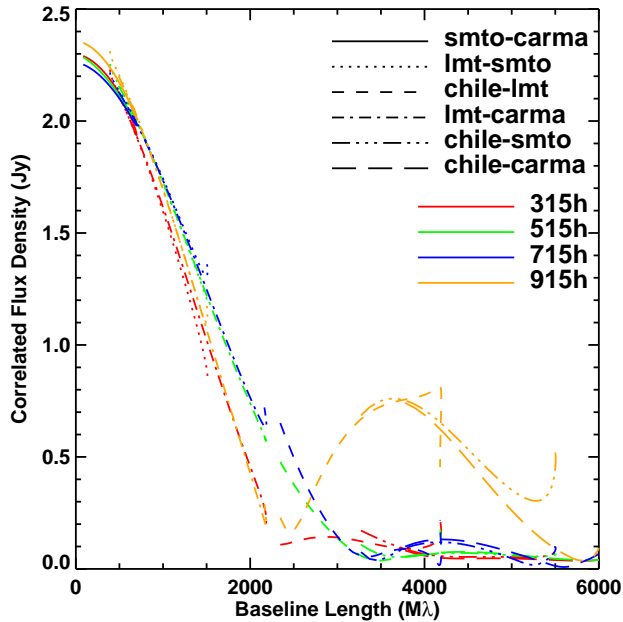}
\caption{\label{vis1dinterp}Visibility amplitude vs. baseline length
  for the best-fitting models interpolated to the uv locations of current
  and future baselines. The local minimum in the visibility amplitudes
for 515h and 715h near $3000 M\lambda$ is a signature of the black
hole shadow.}
\end{figure}

\begin{figure*}
\includegraphics[scale=.9]{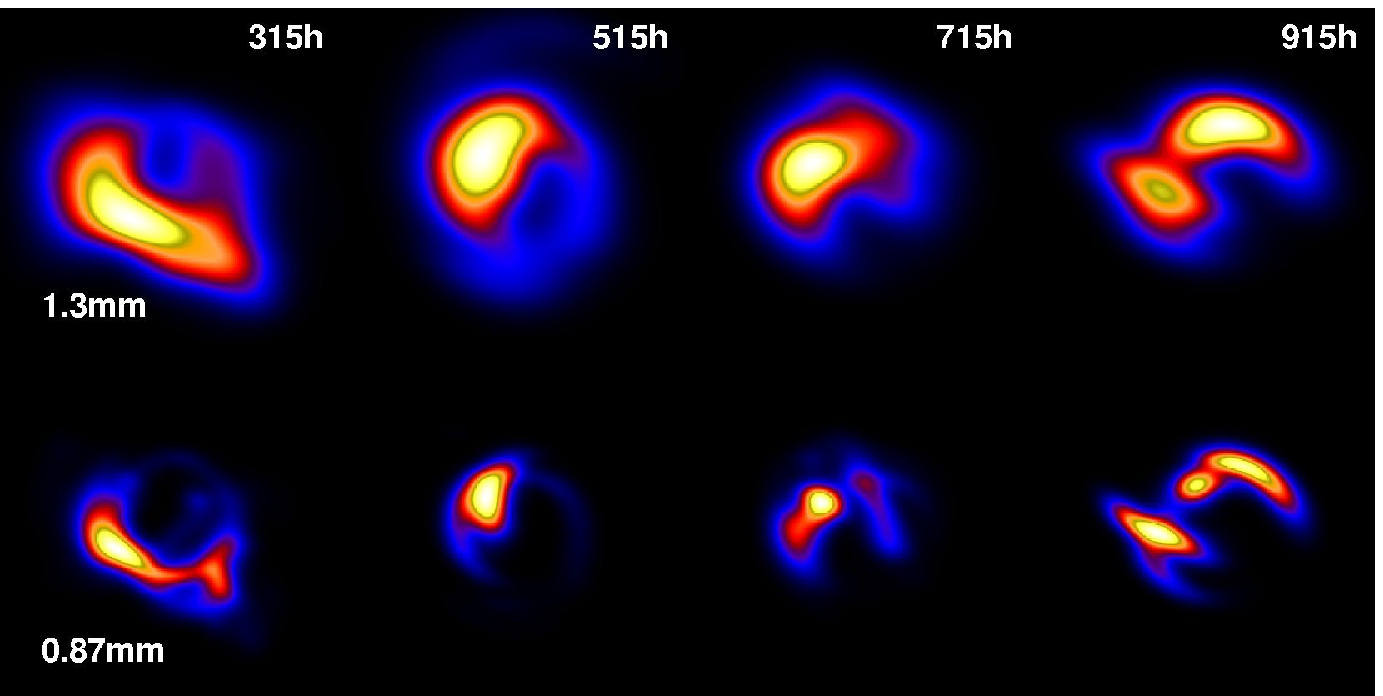}
\caption{\label{dint345}Scatter-broadened best-fitting images at
  $1.3$mm (top) and $0.87$mm (bottom). The scaling is the same as in
  Figure \ref{titeimg}. At $0.87$mm, the images are easier
  to distinguish and the black hole shadow is less prominent.}
\end{figure*}

\subsection{Sgr A* millimetre emission region}

The mm emission in Sgr A* is at the peak of the SED and the transition from optically
thick to thin, and is primarily produced by thermal electrons. The
emission comes from a photosphere where the optical depth $\tau \simeq
1$, so that the temperature of the emitting electrons is given by the
brightness temperature, $T_b$:

\begin{align}
\label{brightnesst}
k T_b =& \frac{c^2 I_{\nu_0}}{2\nu_0^2} \nonumber\\
\simeq& \hspace{2pt}6\times10^{10} \left(\frac{F_{\nu_0}}{3 \hspace{2pt}\rm Jy}\right)
\left(\frac{\nu_0}{230 \hspace{2pt}\rm GHz}\right)^{-2} 
\left(\frac{\Delta\theta}{40 \hspace{2pt}\mu \rm as}\right)^{-2} \rm
K,
\end{align}

\noindent where $F_{\nu_0}$ ($I_{\nu_0}$) is the observed flux
(specific intensity) at frequency
$\nu_0$, and $\Delta \theta$ is the observed linear angular size of
the emitting region scaled to the value found recently from mm-VLBI
observations \citep{doeleman2008,fishetal2011}.  This is effectively a
lower limit for the models, since in the tilted 
disc models the accretion flow can be completely optically thin at
$1.3$mm. However, the electron temperatures in these cases are still 
within a factor of $2$ of $T_b$. This is why even for a wide range
of simulations, $T_i/T_e$, and $\dot{M}$, the electron temperature
is well constrained for both the untilted \citep{dexteretal2010} and
tilted (\S \ref{fitting}) models.

In untilted simulations, the accretion flow effectively consists of
one population of electrons -- those in the inner radii near the
midplane, where temperatures, field strengths, and particle densities
are highest. The best-fitting value of $T_i/T_e$ in untilted simulations
is then that for which this electron population has $T_e \simeq
T_b$. At higher values of $T_i/T_e$, a larger accretion rate is
required to match the observed flux ($j_\nu \propto \dot{M}^2 T_e^2 /
M^3$). The absorption coefficient is less strongly dependent on $T_e$
than the emission coefficient 
($\alpha_\nu \propto j_\nu / B_\nu \propto j_\nu / T_e$ in LTE), so that
absorption becomes increasingly important for higher $T_i/T_e$. The
accretion flow then develops a photosphere which grows outwards in
radius with increasing $T_i/T_e$. This effect is shown in the top row of
Figure \ref{titeimg} for the 90h simulation. Other untilted simulations behave
similarly, except that in conservative simulations \citep[e.g., MBD or
MBQ in][from \citealt{mckinneyblandford2009}]{dexteretal2010} the $T_i$ values are larger and hence so are 
the favored values of $T_i/T_e$. 

The brightness temperature argument holds for the tilted
simulations as well (see \S\ref{fitting}). However, heating
from the standing shocks effectively leads to additional  
electron populations at varying densities, field strengths, and ion 
temperatures, depending on latitude. The densest material with the
lowest ion temperature is near the (tilted) midplane, and is brightest 
for small $T_i/T_e$. The densities and field strengths 
decrease away from the midplane, and these populations of electrons
become more important at higher $T_i/T_e$. The material closer to the
midplane becomes optically thick as $T_i/T_e$ 
increases, but not nearly as much as in untilted simulations. This is
because the emissivity of hotter electrons also increases, and these
electrons are largely unobscured by optically thick material.

Changing $T_i/T_e$ then can either lead to a growing
photosphere, or it can simply increase the emissivity of electrons
that previously had $T_e > T_b$, causing structural changes without
the image necessarily becoming optically thick. This effect can lead
to a complicated dependence of the image on $T_i/T_e$, 
as shown in the bottom two rows of Figure \ref{titeimg} for the 315h
and 515h simulations. While the 315h image becomes increasingly
optically thick with increasing $T_i/T_e$, it does so much more slowly
than the untilted simulations. The 515h simulation begins to grow a
photosphere until hotter electrons dominate the image at $T_i/T_e
\gtrsim 3$, after which the photosphere begins to recede. These images 
remain qualitatively similar from $T_i/T_e \approx 20-200$.

\section{Sgr A* Parameter Constraints}
\label{fitting}

Even with the limited coverage and sensitivity of current mm-VLBI data, it is
possible to estimate parameters of both the black hole and accretion
flow in Sgr A* in the context of either RIAF
\citep{broderick2009,brodericketal2011} or GRMHD models
\citep{dexteretal2010,dexter2011}. The inclination and position angles
are particularly well constrained, and are in excellent agreement
between the two types of models. Both models
assume alignment between the disc and black hole, which is almost
certainly not the case for Sgr A*. 

For the tilted disc simulations with a misalignment of $15^\circ$,
there are many possible models which provide excellent fits to current
mm observations (reduced $\chi^2$ from a single epoch of mm-VLBI of
$\lesssim 0.6$). The wide range of images is mostly due to the strong
changes with observer time and azimuth, both of which have minimal
effects on the structure of untilted disc images. As a result, the
parameters of the model are essentially unconstrained with current
observations (Figures \ref{probdiststilti} and
\ref{probtitemdot}). We can formally estimate the parameter values as: 
$i={80^\circ}_{-42^\circ}^{+10^\circ}$, $\xi={-78^\circ}_{-11^\circ}^{+130^\circ}$,
$\dot{M}={165}_{-161}^{+284} \times 10^{-9} M_\odot \rm yr^{-1}$, and
$T_e={4}_{-1}^{+7}\times10^{10} \rm K$, all to 90 per cent confidence. All
values of $\phi_0$ are allowed at this confidence. More data and a
more complete set of simulations will be required to meaningfully
constrain the parameters. All previous models have assumed a disc tilt
of zero, and have found much stronger constraints on Sgr A* parameters
than we find for a tilt of $15^\circ$. Previously reported parameter
estimates are then invalid except for either very low values of tilt
or spin ($a < 0.3$ or $\beta < 15^\circ$ are the upper limits from
these simulations). 

As in untilted models, face-on inclinations are disfavored, primarily
because they are too optically thin to explain the observed spectral
indices \citep{moscibrodzka2009}. A wide range of position angles are
viable, but the location of the broad peak in probability density
corresponds to the range found for untilted models
\citep{dexteretal2010,brodericketal2011}. In each model, an observer
azimuth is favored where velocities 
downstream of one of the non-axisymmetric shocks are approaching the
observer. For 915h, both such orientations are favored, while in
the other simulations the probability distribution is asymmetric with
one standing shock favored.

As discussed above (\S\ref{modeling}), it is non-trivial
to find all viable values of $T_i/T_e$ for the tilted
simulations. Since a higher accretion rate is required at higher
$T_i/T_e$ to match the observed flux of Sgr A*, this leads to a wide
range of possible accretion rates, unlike the untilted case. However,
the electron temperature is still well constrained. Lowering $T_i/T_e$
and increasing the accretion rate serves to pick out different
electrons, but does not change their temperature by more than a factor
of two.

\begin{table}
\caption{\label{bfit}Best Fit Model Parameters}
\begin{scriptsize}
\begin{center}
\begin{tabular}{lcccccc}
        \tableline
	\tableline
 Name & Spin & $\dot{M} (10^{-9} M_\odot \mathrm{yr}^{-1})$ & $i$
 & $\xi$ & $\phi_0$ & $T_i/T_e$\\
        \tableline
315h & 0.3 & $130$ & $70^\circ$ & $-13^\circ$ & $\pi/4$ & 3 \\
515h & 0.5 & $18$ & $40^\circ$ & $-81^\circ$ & $-\pi/4$ & 1\\
715h & 0.7 & $10$ & $90^\circ$ & $-70^\circ$ & $-\pi/2$ & 1\\
915h & 0.9 & $6$ & $80^\circ$ & $-38^\circ$ & $3 \pi / 4$ & 1\\
	\tableline
\end{tabular}
\end{center}
\end{scriptsize}
\end{table}

\section{Model Properties}

Despite the lack of robust or meaningful parameter constraints, there
are many generic properties of the images, spectra, and light curves
of tilted accretion disc models 
with important observational implications for Sgr A*. The best-fitting
models from each simulation are representative of the range of
viable possibilities from our fitting, and in this section these
models are studied in detail. Their parameters are given in Table
\ref{bfit}. 

\label{sec:model-properties}
\subsection{Images}
\label{sec:images}

The importance of non-axisymmetric standing shocks for Sgr A* images
is shown in Figure \ref{shockimgs}. Images from a
tilted simulation (715h, left panel) are dominated by material heated
by the standing shocks, which appear as non-axisymmetric, roughly $m=2$
structures. Any non-axisymmetric structure in untilted
images (right panel of Figure \ref{shockimgs}) travels at the orbital
speed, while the standing shocks move at the much slower precession
speed. In addition to the choice of $T_i/T_e$ described above, images from
tilted simulations are then primarily set by observer azimuth, which
determines the viewing orientation of the standing shocks. 

The asymmetry of the tilted disc images leads to a wide range of
complex image morphologies, which depending on the model and observer
time can be dominated either by the post-shock fluid, dense material
near the midplane of the tilted disc, or other hot electrons. This is
the primary reason that the model parameters cannot be well constrained thus
far. Despite the additional complexity, viable images tend to be
similar. This is demonstrated using the best-fitting models from
all simulations, whose $1.3$mm images (top row of Figure
\ref{dint230}) are all ``crescents,'' as found previously for images 
from untilted simulations \citep{dexteretal2010}. The crescent
morphology results from a combination of strong Doppler beaming from orbital disc motion,
which causes asymmetry between approaching and receding material, and
gravitational lensing, which causes the back of the accretion flow to
appear above and below the black hole in the image. The crescent
morphology is especially apparent when the images are
convolved with interstellar scattering (second row of Figure
\ref{dint230}), which at $1.3$mm blurs 
most of the small scale image structure. The black hole shadow is
apparent as a central minimum in all images, at the orientation
corresponding to light 
coming from the back of the accretion flow, which is bent above and
below the black hole. The shadow appears in the 
visibility amplitude (third row of Figure \ref{dint230}) as a local
minimum on the same orientation at a baseline whose length corresponds
to the size scale of the shadow in the image. It also appears
at the same location as a region of rapidly varying visibility phase
(fourth row of Figure \ref{dint230}). 

We can make predictions for future mm-VLBI observations by
interpolating visibilities to the locations of future
baselines between ALMA/APEX in Chile, LMT in Mexico, and current
telescopes: SMTO in Arizona, JCMT/SMA in Hawaii, and CARMA in
California. The results are shown in Figure \ref{vis1dinterp} for the
visibility amplitudes for the best-fitting models. The models make a
wide variety of predictions for future baselines, and the
black hole shadow appears as a local minimum in the 
visibility amplitude near $3000M\lambda$ in the 515h and 715h models. 

In addition to visibility amplitudes, phases also carry important 
information about the image structure. The closure phase is formed
by summing phases over a triangle of baselines, and is more robust to
instrumental errors than individual phase measurements. For favored
(and most) position angles for all best-fitting models, the closure phase
is near zero, in agreement with recent mm-VLBI observations
\citep{fishetal2011}. In addition, the black hole shadow shows up as a
highly variable closure phase either vs. position angle or time for
position angles where the shadow is accessible. For nearly co-linear baselines
such as those including current mm-VLBI telescopes, the closure phase
tends to be near zero except for viewing orientations corresponding to
the black hole shadow. However, phase information is more interesting
for other baseline triangles. Depending on
viewing orientation and model, a wide range of possible closure phases
are possible, many of which differ significantly from zero. Measuring
closure phase in future mm-VLBI campaigns will therefore provide
significant additional constraints on the models. 

At $0.87$mm (Figure \ref{dint345}), the other potential wavelength for
future mm-VLBI observations, the effects of interstellar scattering are greatly
reduced. These images have intrinsically steeper emissivities,
however, so that the emission is more concentrated and the black hole
shadow is less apparent. For these reasons, shorter wavelengths may be
better suited for constraining parameters and testing accretion models
than for detecting the black hole shadow. The structure and total flux
of the accretion flow at this wavelength are poorly constrained, and
the different simulations predict a wide range of both possible
image structures and fluxes. For this reason, $0.87$mm VLBI may be 
better for constraining the models. Even a simultaneous spectral 
constraint at $0.87$mm along with $1.3$mm could distinguish between
many of the possibilities. 

\subsubsection{Implications for mm-VLBI observations}

\begin{figure*}
\includegraphics[scale=.8]{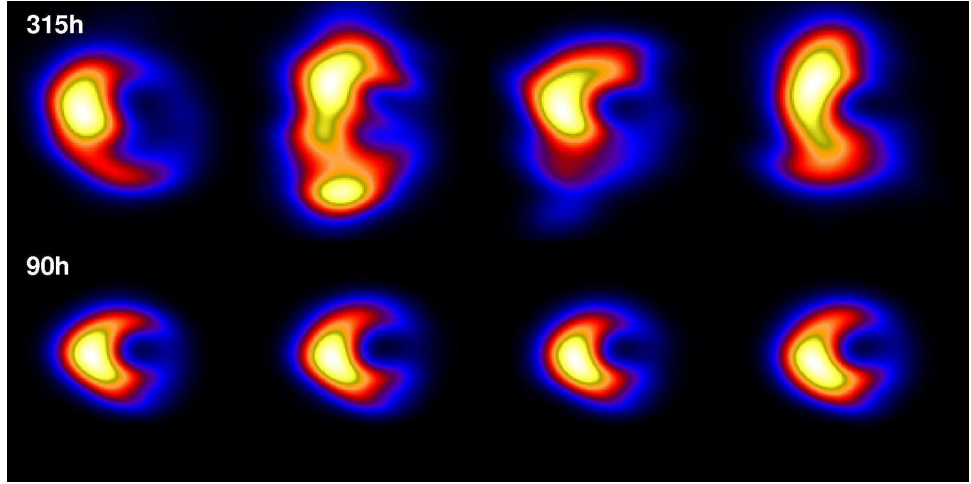}
\caption{\label{tiltuntiltcompare}Images at four observer times for
  the 315h (top row) and 90h (bottom row) simulations including the effects of interstellar
  scattering. The scaling is the same as in Figure \ref{titeimg}. The structure of images from untilted models
  such as 90h show very little time variability, while those from
  tilted models can change significantly on hour timescales.}
\end{figure*}

While a crescent morphology is often found for mm tilted disc images
of Sgr A* (especially at $1.3$mm), similar to previous models of untilted discs, their
time-dependent structures are completely different. Untilted disc
images are basically static, as shown in the bottom row of Figure
\ref{tiltuntiltcompare}. Four images from the 90h simulation are shown
spanning the simulation duration ($\simeq 25$ hours for Sgr A*). Even as the flux
varies by $\simeq 50$ per cent, the image structure remains virtually
unchanged. Tilted disc images, on the other hand, can undergo major
structural changes even on relatively short timescales (hours, top row
of Figure \ref{tiltuntiltcompare}). Thus far, no significant 
structural changes have been detected with mm-VLBI, consistent with
the untilted models. With better coverage, higher sensitivity, and more
epochs of data, time variations in structure will become a powerful
means to distinguish between untilted and tilted accretion disc
models.

\begin{figure*}
\includegraphics[scale=.8]{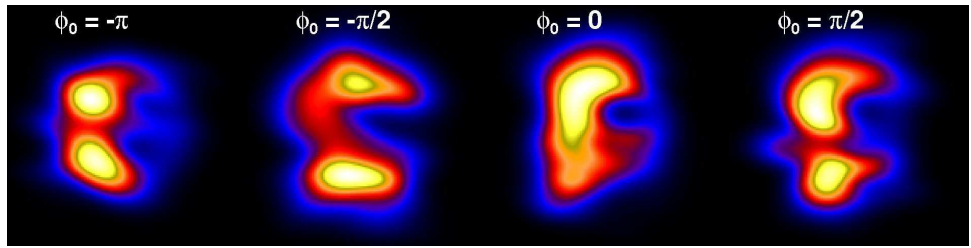}
\caption{\label{phiobsimgs}Images at four observer azimuths for
  the 315h simulation including the effects of interstellar
  scattering. The scaling is the same as in Figure \ref{titeimg}. The image structure changes significantly with observer
  azimuth, and these changes will be observable on the precession timescale.}
\end{figure*}

\begin{figure}
\includegraphics[scale=.8]{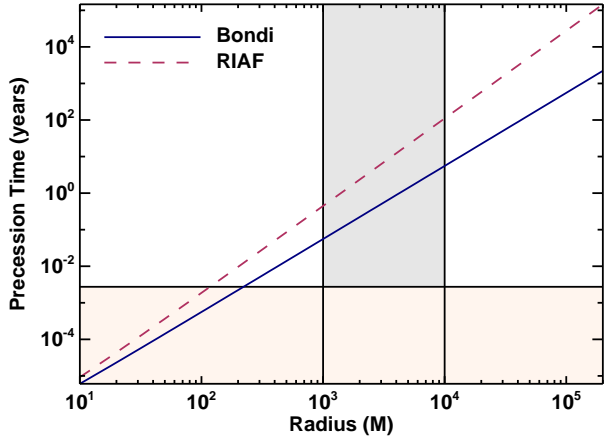}
\caption{\label{tprec}Precession time as a function of outer accretion
  flow radius ranging from the event horizon to the Bondi radius for
  radial power law surface density profiles corresponding to RIAF and spherical (Bondi)
  accretion models. If the effective outer accretion flow radius is
  $\lesssim 10^{3-4} M$ \citep[gray shaded region,][]{cuadraetal2006}, then precession 
  should occur on year timescales, potentially observable as changes
  between epochs of mm-VLBI observations. Only for extremely small
  outer radii would significant precession occur within a single epoch of
  observations (day timescales, yellow shaded region).}
\end{figure}

Lense-Thirring precession (\S\ref{sec:tilted-black-hole}) is another
source of possible inter-epoch variability for tilted disc models of
Sgr A*. The image structure depends strongly on the observer
azimuth (Figure \ref{phiobsimgs}), which would effectively change as
the disc precesses. Future mm-VLBI observations could then find
evidence for this general 
relativistic effect in the Galactic centre. Unfortunately, the
relevant timescale is highly uncertain. It depends sensitively
on the effective ``outer'' radius of the accretion flow, its radial
density profile, and the black hole spin. The outer radius is
especially problematic, since both its location and interpretation are
uncertain for the collisionless plasma around Sgr A*. In terms of
these parameters, the precession time is given by \citep{fragile2007}: 

\begin{equation}
\label{eq:tprec}
t_{\rm prec} \simeq \frac{1.3}{a} \left(\frac{r_{\rm out}}{1000M}\right)^2 \rm yr,
\end{equation}

\noindent where $r_{\rm out}$ is the effective outer radius. We
have assumed $\rho \propto r^{-3/2}$ as expected for Bondi accretion
and as found by hydrodynamical simulations of large-scale accretion
onto Sgr A* \citep{cuadraetal2006}. The precession time is shown in
Figure \ref{tprec} as a function
of outer radius for both the Bondi and RIAF solutions. 
In both cases, for large ranges of outer radius, the precession
timescale is likely to be longer than the length of individual mm-VLBI campaigns (days,
shaded gray region in Figure \ref{tprec}) but
shorter than the length of the entire Event Horizon Telescope experiment
(years). This is the case, for example, for both density profiles if
the effective outer radius corresponds to the circularization radius
found for the accretion flow by \citet{cuadraetal2006} (orange shaded
region in Figure \ref{tprec}). Alternatively, if the outer radius is extremely small ($r_{\rm max} \lesssim
100 M$), as \citet{pangetal2011} found using MHD simulations with a
strong large-scale magnetic field, then it could be possible for 
individual mm-VLBI epochs to span a precession time. 

Current data are insufficient to meaningfully test for changes in
observer azimuth (and precession) between or within epochs. The
prospects for detecting Lense-Thirring 
precession with Event Horizon Telescope observations will be
considered in more detail in future work.  

\begin{figure*}
\includegraphics[scale=.7]{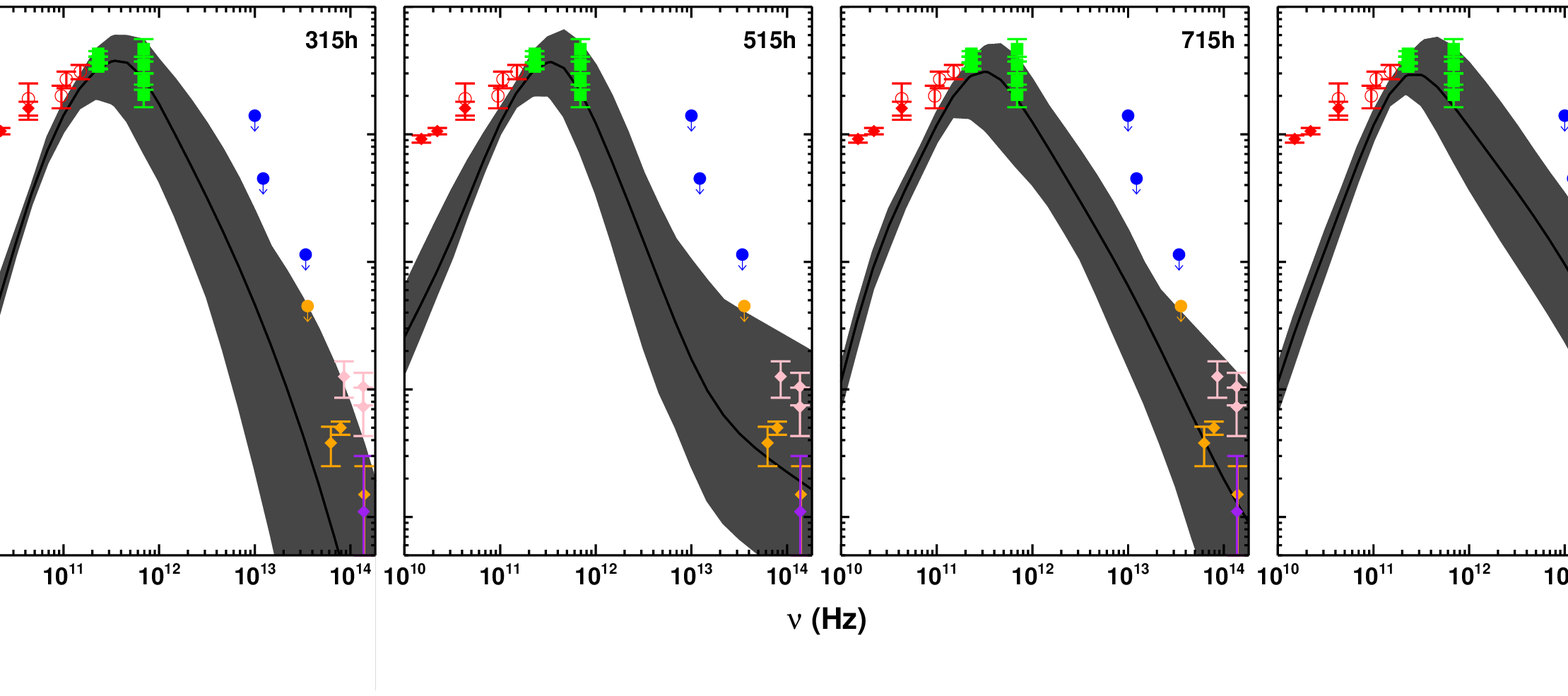}
\caption{\label{bfitspectra}Best-fitting model spectra compared to Sgr A* data. The
  solid curves are the median values at each frequency over observer
  time, while the dark gray envelope shows the range. Radio data are
  from \citet[][open circles]{falcke1998} and \citet[][filled
  diamonds]{an2005}. The models are fit to the mm data \citep[filled
  squares,][]{marronephd}, but are in good agreement with the
  mean/median \citep[orange and
  purple,][respectively]{schoedeletal2011,doddsedenetal2011} and
  flaring \citep[pink,][]{genzel2003} NIR data with no additional free
  parameters. They also satisfy observational upper limits in the
  mid-IR \citep[blue upper limits,][and references
  therein]{meliafalcke2001}. At lower frequencies the non-thermal
  emission is produced outside of the simulation domain, either in the
  accretion flow \citep{yuanquataert2003} or in a short jet
  \citep{falckemarkoff2000}.}
\end{figure*}

\subsection{Spectra}
\label{sec:spectra}

Spectra from the best-fitting tilted disc models are shown in Figure
\ref{bfitspectra}. The black lines show the average over
observer time, while the dark gray envelope indicates the range. These
models cannot explain the radio emission outside the 
sub-mm bump due to their limited spatial extent and lack of
non-thermal emission. The IR tilted disc spectra, however, are much
different than any from previous MHD simulations. With purely thermal 
electrons and no additional free parameters, all best-fitting models
to the mm data are consistent with upper limits in the far- and mid-IR 
while producing the observed flux range in the near-IR (NIR) \citep[although the 915h model 
marginally violates the mid-IR upper limit from][]{schoedeletal2011}. 

The mean spectra from the 
515h, 715h, and 915h models are in remarkable agreement with reported mean values
from \citet{schoedeletal2011} and the median value from
\citet{doddsedenetal2011}. Further, the NIR spectral indices 
(defined as $F_\nu \propto \nu^{-\alpha_{\rm NIR}}$ and listed in
Table \ref{irvar}) from these models are within the range reported for 
Sgr A* \citep{ghezetal2005,gillessenetal2006,hornsteinetal2007}. The
spectral index from the 515h model 
is in remarkable agreement with the result found by
\citet{hornsteinetal2007}, while those from the 715h and 915h models
agree with the redder reported spectral index of \citet{gillessenetal2006}. 
The 315h spectrum underproduces the average 
NIR emission and its spectral index is steeper than observed. Other model
spectra from these simulations may provide excellent fits to the
observed NIR emission in Sgr A* as well, and the NIR data could be
included in the joint fitting in future studies.

\begin{figure*}
\includegraphics[scale=1]{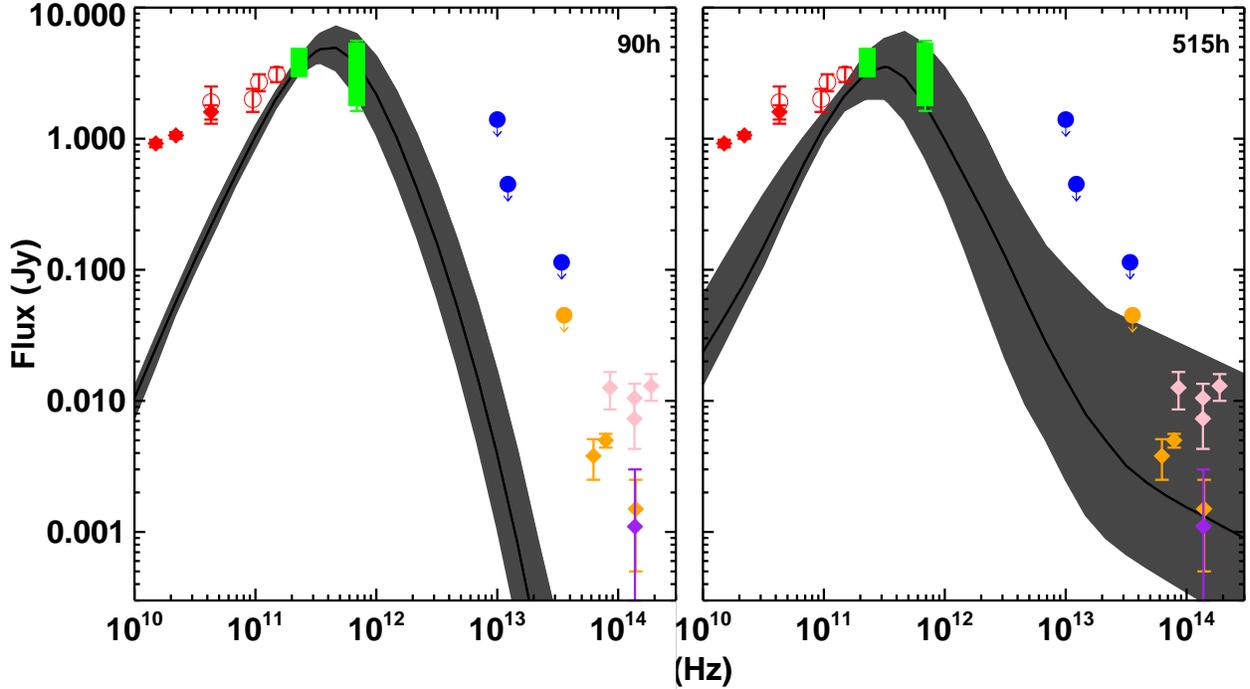}
\caption{\label{tiltuntiltspectra}Spectra from 
  best-fitting untilted (90h, left) and tilted (515h, right) 
  simulations. Sgr A* data are the same as in Figure
  \ref{bfitspectra}. In both cases the spectra are fit to the green sub-mm
  points. Multiple electron populations from shock heating in tilted
  discs can naturally produce the observed NIR emission, which is
  underproduced by $\approx 2$ orders of magnitude in untilted
  simulations. Other untilted simulations exhibit larger amplitude
  variability than 90h, but its spectral shape is representative.}
\end{figure*}

Spectra from untilted disc simulations have similar shapes to that
shown for 90h in the left panel of Figure 
\ref{tiltuntiltspectra} \citep[e.g.,][]{moscibrodzka2009}. The NIR
emission is on the exponential tail of the synchrotron spectrum. For
this reason, the untilted models underproduce the observed emission
by several orders of magnitude and have much redder spectral indices
than observed. Comparing the 90h spectrum with that from 515h (right panel
of Figure \ref{tiltuntiltspectra}) shows that distinct electron
populations (and/or a non-thermal distribution function) are crucial for
simultaneously producing the mm to NIR spectrum. 

The NIR emission in the tilted models is produced by extremely hot electrons
($T_e \approx 1-4 \times10^{12}$ K) near the black hole ($r \approx 3
M$). Heating from the standing shocks 
in the tilted disc models allow for multiple electron populations,
some of which can be at very high temperatures. These can effectively
mimic a power law tail to the synchrotron emission, especially since
they are at significantly lower particle densities than the electrons
producing the mm bump (\S \ref{sec:physical-origin-mm}). This physical
heating mimics the effect 
of adding a non-thermal tail to the electron distribution function
\citep[e.g.,][]{yuanquataert2003}. 

These hot electrons are also the reason we discard models with 
$T_i/T_e \gtrsim 3$ for the 515h and 715h simulations, despite the
fact that some of them can fit the mm-VLBI and spectral
observations. Larger accretion rates are required to match the
observed mm flux for larger $T_i/T_e$. The peak synchrotron
frequency scales as,

\begin{equation}
 \nu_c \sim B T_e^2 \sim \sqrt{\dot{M}} T_e^2,
\end{equation}

\noindent while the peak flux scales roughly as 

\begin{equation}
F_{\nu_c} \sim n B^2T_e^2 \sim \dot{M}^2 T_e^2,
\end{equation}

\noindent so that $\nu_c$ decreases for larger
$T_i/T_e$ at fixed flux. This scaling causes the high frequency parts of the spectra to
shift up in normalization and down in frequency. The approximately power law tail
can extend into the optical/UV when $T_i/T_e=1$, and increasing its
normalization leads to the violation of many IR upper limits. Although
we do not include IR data in the fitting, we do discard models
which violate many IR upper limits.

\begin{table}
\caption{NIR Variability Parameters\label{irvar}}
\begin{minipage}{6cm}
\begin{small}
\begin{center}
\begin{tabular}{lccc}

        \tableline
        $$ & $\alpha_{\rm NIR}$ & $\exp{\mu}$ (mJy) & $\sigma$\\
        \tableline
315h & $-4.1 \pm 1.0$ & $(0.7-1.9)\times10^{-2}$ & $11-18$\\
515h & $-0.5 \pm 0.1$ & $1.1-1.3$ & $1.8-2.0$\\ 
715h & $-1.6 \pm 0.7$ & $0.5-0.7$ & $3.0-3.4$\\ 
915h & $-2.2 \pm 0.4$ & $1.0-1.3$ & $1.6-1.7$\\
	\tableline
\end{tabular}
\end{center}
\end{small}
\end{minipage}
\end{table}

\subsection{Millimetre to infrared variability}

\begin{figure*}
\includegraphics[scale=.7]{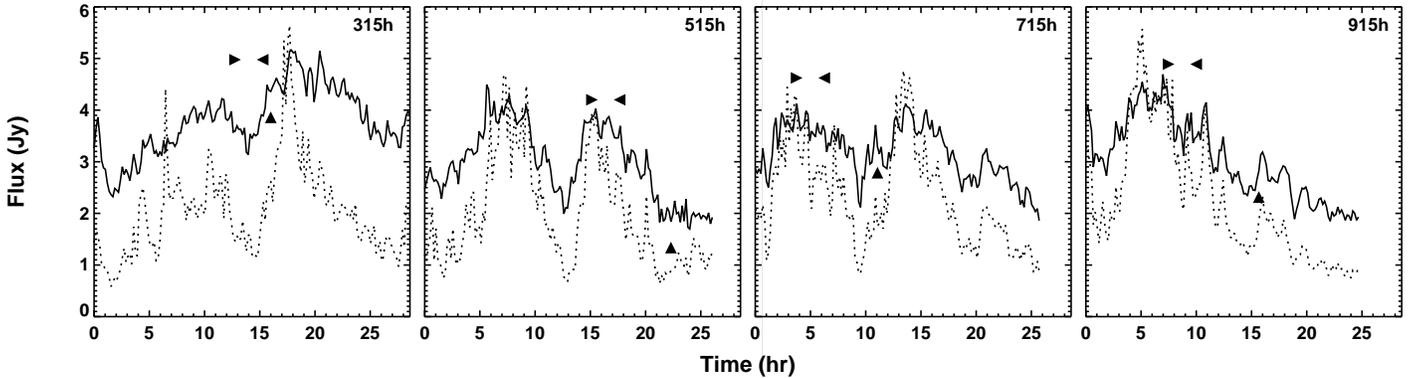}
\caption{\label{lcurves}Light curves from the best-fitting models at
  $1.3$mm ($230$ GHz, solid lines) and $0.4$mm ($690$ GHz, dotted lines). The 
  times of best fit to the visibility data are shown as upward
  triangles, while the ranges of times averaged for the best fit to
  the spectral data are bracketed by left and right facing
  arrows. The simulated light curves exhibit 
  $50$ per cent variability on hour timescales, similar to the untilted
  case and in qualitative agreement with observations. The mm spectral
index increases with flux, because the optical depth is lower at
higher frequency.}
\end{figure*}

Radiative models of Sgr A* based on simulations are thus far the only
to self-consistently include variability information
\citep{goldston2005,dexter2009,dexteretal2010,dolenceetal2012}. Best-fitting
models to mm-VLBI and spectral information from untilted GRMHD 
simulations naturally produce millimeter variability qualitatively
consistent with that 
observed \citep{dexter2009,dexteretal2010}. This variability is due to
relatively global ($m=0$, $1$) fluctuations in the particle density and
magnetic field strength, and is strongly correlated with the accretion
rate. We find similar mm light curves for the best-fitting tilted disc
models, as shown in Figure \ref{lcurves}. All of the best-fitting
light curves exhibit variability on hour timescales with $\lesssim
50$ per cent amplitudes, roughly in agreement with 
observations. Millimeter ``flares'' would be identified as two per
light curve, consistent with their observed frequency in Sgr
A*. However, tilted disc models produce larger amplitude and longer
duration flares than their untilted counterparts \citep[cf. Figure 6
of][]{dexteretal2010}. The variability is still strongly correlated
with the accretion rate, which also varies more in tilted disc
simulations. 

\begin{figure*}
\includegraphics[scale=.7]{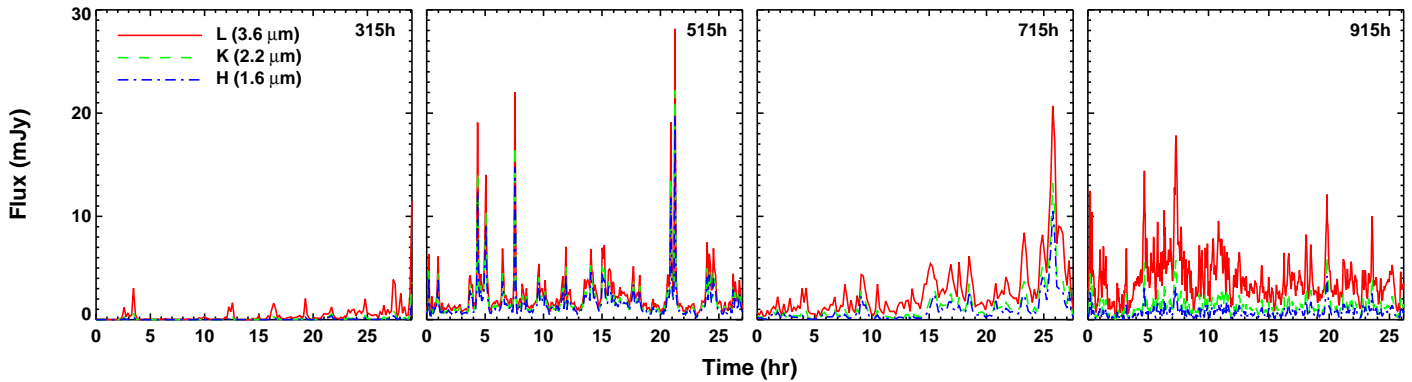}
\caption{\label{lcurveshkl}Near-infrared light curves from the
  best-fitting models in the H, K, and L bands. All of the simulated NIR light curves are highly variable on
  short ($\lesssim \rm hr$ timescales). The 315h simulation
  underproduces the observed NIR flux and its variability. The 515h, 
  715h, and 915h models produce the observed flux range and exhibit
  ``flares'' with roughly the right amplitudes (factors $\lesssim 30$)
  and timescales ($\simeq 1 \rm hr$). The 315h
  and 715h spectral slopes are time-variable, while those from the
  515h and 915h models are nearly constant with NIR flux.}
\end{figure*}

As discussed above, tilted disc models naturally produce roughly the
correct mean levels of IR emission, very different from previous
numerical accretion flow models based solely on thermal electrons
\citep[e.g.,][]{goldston2005,moscibrodzka2009}. The H 
($1.6\mu$m), K ($2.2\mu$m), and L ($3.6 \mu$m) band light curves 
from the best-fitting models to the mm data are shown in Figure
\ref{lcurveshkl}. In all cases, the NIR emission is highly variable,
with factor of $\simeq 5-20$ increases over the mean level on
$\lesssim 1$hr timescales, consistent with observations. The time
sampling in the light curves is $1-10$ minutes, depending on 
the simulation. The 915h simulation, with 1 minute sampling, easily
resolves the NIR flux variations. The shortest significant flux
variations (factors of $\gtrsim 2$) occur on
timescales $\simeq 5-15$ minutes. Substructure in the 315h and 715h
light curves, with $10$ minute sampling, is only marginally resolved. 

The various models differ substantially in the
predicted NIR spectral slope (see \S and Table
\ref{irvar}). Similarly, two of the spectral slopes 
are time variable (315h and 715h),
while the others are nearly constant (515h and 915h). The flux
distributions of all of the light curves are fairly well described by a
log-normal:

\begin{equation}
P(F_\nu; \mu, \sigma) = \frac{1}{\sqrt{2\pi}\sigma F_\nu} \exp\left(-\frac{(\ln F_\nu -
    \mu)^2}{2 \sigma^2}\right),
\end{equation}

\noindent with parameters $\mu$ and $\sigma$. The median is
$\exp(\mu)$, and $\sigma$ represents a multiplicative standard
deviation. This model provides a better description than a standard
Gaussian, as \citet{doddsedenetal2011} found for Sgr A* data. The
parameters inferred from our K-band light curves are given in Table
\ref{irvar}. The ranges represent the results of fitting with various
bin sizes.

\begin{figure}
\includegraphics[scale=.75]{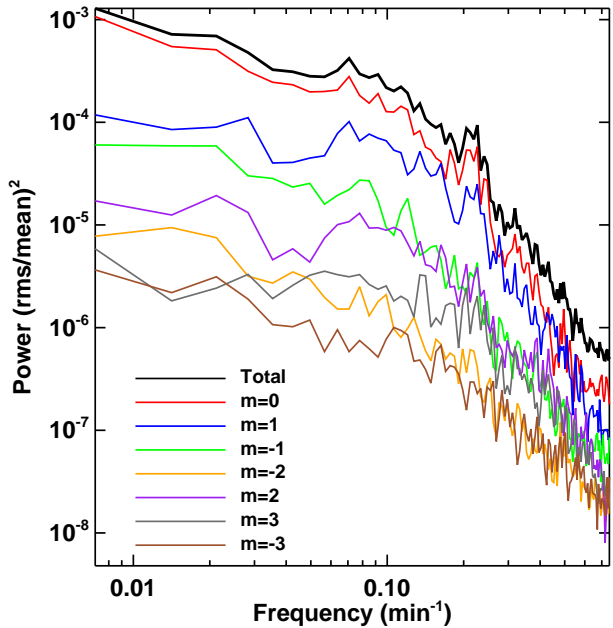}
\caption{\label{psdm}NIR K-band ($2.2\mu$m) power spectrum (thick
  solid curve) from the best-fitting 915h model, averaged over $10$
  evenly-spaced light curve 
  segments. The power spectrum is further decomposed in azimuthal
  mode, as done in \citet{dolenceetal2012}. Both the overall power
  spectrum and those from individual azimtuhal modes are well
  described by broken power laws with break frequencies $\sim
  10 \rm min^{-1}$. The full power spectrum has slopes $\simeq -1$ and
  $-4.5$.}
\end{figure}

Power spectra from all the tilted disc models are qualitatively
similar to that for 915h (thick solid curve in Figure \ref{psdm}). The
slope is fairly shallow (slope $\simeq -1$) on timescales
significantly longer than that 
of the marginally stable orbit ($\gtrsim 10$ min), although the
effective inner disc edge is well outside of this location
\citep{fragiletilt2009,dexterfragile2011}. On shorter timescales the
power spectrum steepens to a slope $\simeq -4.5$. Fit with a single power
law, the slope is $\simeq -2.3$, consistent with observations
\citep{meyeretal2008,doetal2009}. 

\citet{dolenceetal2012} found a broken power law power spectrum in a
model of Sgr A* from an untilted GRMHD simulation with a similar break
frequency. However, we find no significant quasi-periodic oscillations
similar to what they reported near the frequency of the marginally
stable orbit. Using many cameras spaced in observer azimuth, they
decomposed the power spectrum into azimuthal modes (their Eq. 1 and
Fig. 1). A comparison is shown in Figure \ref{psdm}. Unlike in their
models, the $m=0$ mode is dominant, with $m=1$, $2$ only contributing
on the shortest timescales. These discrepancies could be due to
differences in the simulations (either numerical algorithms or
duration), or due to physical differences between tilted and untilted
accretion discs.

\subsection{Physical origin of time-variable mm and NIR emission} 
\label{sec:physical-origin-mm}

The mm emission in the tilted disc models arises from hot, dense
electrons in the inner radii of the simulations. Typical parameters of
mm electrons are similar to those in untilted discs: $n \sim 10^{7}
\rm cm^{-3}$, $B \sim 50 \rm G$, $T_e \sim 6\times10^{10} \rm K$. As
discussed above, this tends to be where densities and field strengths
are highest: either in the post-shock fluid or in
the midplane of the inner radii of the tilted accretion flow. Due to
the asymmetry in the shocks and subsequent infall, the emission geometry is 
different from untilted simulations. The mm variability in the tilted
disc models is, however, essentially
the same as in untilted models. The variability is caused by 
relatively global fluctuations in the particle density and magnetic field strength. Most
of the power in these fluid variables as well as in the emissivity is
in the $m=0$ mode, although there is also significant power in
$m=1$, $2$, particularly in the form of spiral waves. The $0.4$mm (690
GHz) light curve is strongly correlated with the accretion rate
through the inner boundary, suggesting that at least the $m=0$ power
is from global changes in the accretion rate. The global nature of the
variability may explain its similarities in tilted and untilted
models, despite significant differences in their dynamics. 

\citet{dexteretal2010} found that a simple approximate emissivity,
$j_\nu \propto n B^2$, captured the shape of the mm light curves from untilted
simulations of Sgr A*. This is the expected form for isothermal
synchrotron emission with $T_e \simeq 6\times10^{10}$K, and
demonstrates that temperature fluctuations are not important for
driving the variability. We find the same result for the tilted disc
models. This variability mechanism is not magnetic reconnection, since
the tilted simulations do not capture heat lost by grid-scale magnetic
dissipation. It is also not orbiting hot spots, as the $m=0$ power
dominates the fluctuations. 

The NIR emission in the models is entirely optically thin, and
dominated by very small regions at small radius ($r \simeq 3\rm M$) in
the simulation with lower densities 
($n \sim 10^5 \rm cm^{-3}$), similar magnetic field strengths ($B \sim
50 \rm G$), and hot temperatures ($T_e \gtrsim 10^{12}
\rm K$). This location is fairly insensitive to spin, and is inside
the marginally stable orbit for all but the 915h simulation. These are
lower particle densities and higher electron 
temperatures than previously reported for IR-emitting electrons from
axisymmetric, untilted GRMHD simulations \citep{moscibrodzka2009}. In
their simulations, the NIR emission was dominated by heating from numerical magnetic
reconnection in current sheets, which is thought to be an artifact of
axisymmetry. In our model, the NIR-emitting electrons are heated by
the non-axisymmetric standing shocks. This is the only source of
entropy generation in these simulations, since heat from numerical
reconnection is lost from the grid. For example, the maximum 
temperature obtained anywhere in the 90h simulation is $\lesssim
10^{12} \rm K$. 

The NIR ``flares'' in our models are due to fluctuations in the
particle density, field strength, and electron temperature of these
hot post-shock electrons. The 915h K-band light curve can be
recovered approximately with an emissivity $j_\nu \propto n B^2$, as
long as only hot electrons are included ($T_e > 5\times10^{11} \rm
K$).  In the 515h and 715h simulations (with hotter electrons), the emissivity
can be approximated as $j_\nu \propto n B T_e$. Temperature fluctuations
are important in the 515h and 715h simulations, but not in 915h. The
approximate forms can be understood by the variations in emissivity
expected from fluctuations in the individual quantities at the
appropriate ratio of $\nu/\nu_c$. This ratio is near 1 for the hotter
515h and 715h simulations, and $\approx 10-20$ for 915h. Closer to
$\nu=\nu_c$, the emissivity is less strongly dependent on $B$ and
$T_e$. 

\begin{figure*}
\includegraphics[scale=.75]{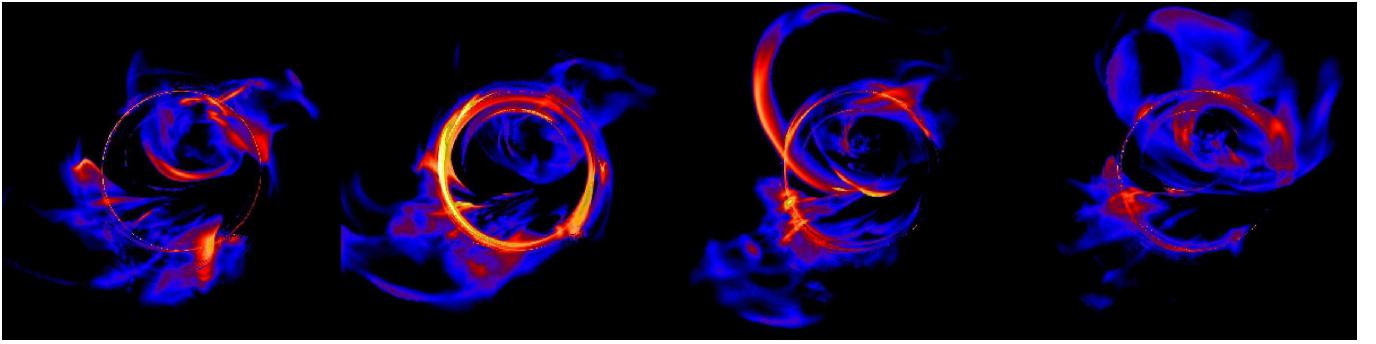}
\caption{\label{515hlens}H-band (1.6$\mu$m) images from the 515h model
  spaced $\simeq 5
  \rm min$ apart during an NIR ``flare'' with a factor of $\simeq 4$ flux variation between
the first and second images. The images are scaled logarithmically
with a dynamic range of 2048, and the panel size is $85\times85 \mu$as. The flare emission is clearly lensed and
concentrated at the photon ring, corresponding to the unstable photon
orbit in the Kerr spacetime.}
\end{figure*}

General relativistic effects are also important for the NIR light
curves from 515h and 715h. Many of the flares in these
models correspond to extreme gravitational lensing events, where
emission from hot electrons behind the black hole is lensed into a
bright ring. An example from the 515h model is shown in Figure
\ref{515hlens}. A flare occurs between the first and second panels
when hot gas behind the black hole is strongly lensed, concentrating
its emission in a bright ring at the circular
photon orbit. In the subsequent frames, this hot material passes by
the black hole and moves to larger radius, leading to a decay of the
flare. The lensing acts as a filter, enhancing the emission from 
electrons at larger radius ($r \gtrsim 3 M$), and can produce
modulations in the light curves of order the observed flux.

The NIR image structure is highly variable, and so is its
centroid. The K-band centroid position is shown in Figure
\ref{ircentroid}. The typical excursions are $\simeq 30-50 \mu$as in
all best-fitting models except that from the 915h simulation. The
position wander is only weakly correlated with the NIR flux. All
periods of rapid variability and very high NIR fluxes also exhibit
significant centroid motions, but not all periods of large position
wander correspond to NIR flares. 

The NIR spectral index is set by the critical frequency for
synchrotron emission, 

\begin{equation}
\nu_c = 6\times10^{12} \left(\frac{T_e}{10^{12} \rm K}\right)^2 \left(\frac{B}{50
    \rm G}\right)^{1/2} \rm Hz.
\end{equation}

\noindent This primarily depends on electron temperature,
which varies between simulations. The NIR electrons are coldest in the
315h simulation, and hottest in the 515h simulation. Somewhat
surprisingly, the shock heating is less effective in the 715h and 915h
simulations. \citet{dexterfragile2011} argued that the
non-axisymmetric standing shocks are stronger in higher spin
simulations. Colder NIR electrons and lower critical
frequencies lead to more negative spectral indices. These also tend to
be more variable, since the emissivity
depends strongly on frequency when $\nu/\nu_c \gg 1$. When $\nu/\nu_c
\simeq 1$, as for the 515h simulation, the emissivity is fairly flat,
and the spectral index is nearly constant. 

The mm and NIR light curves are uncorrelated in this model, since they
arise from different populations of electrons. The full 690 GHz and
K-band light curves and their cross-correlation are shown in Figure
\ref{mmircrosscor}. Although the full duration light curves are
uncorrelated, $6$ hour segments can produce spurious peaks in the
cross-correlation with hour lags, similar to those reported for Sgr A*
\citep[e.g.,][]{yusefzadehetal2008}. 

\section{Discussion}
\label{sec:discussion}

Geometrically thick, low luminosity accretion flows do not align with
their central black holes. The current source of accretion onto Sgr A*
is winds from massive stars, whose net angular momentum is almost
certainly unrelated to
the spin axis of the black hole. Both the current accretion rates onto
Sgr A* at the Bondi radius and in the inner radii are orders of
magnitude smaller than required for sufficient mass accretion to have
changed the black hole spin orientation 
over a Hubble time. Therefore, it is highly unlikely that the black hole
spin axis in Sgr A* aligns with the angular momentum of the accreting gas. 
We have constructed the first radiative models of Sgr A* from
misaligned (tilted) 
black hole accretion flows by performing radiative
transfer calculations on data from GRMHD
simulations (Table \ref{sims}). These models provide
an excellent description of existing observations, and are
significantly different from any theoretical model previously
considered. Even for the modest $15^\circ$ misalignment considered between black
hole spin and accretion angular momentum axes, the (thermo)dynamics of
the accretion flow change drastically. Non-axisymmetric shocks from
the convergence of eccentric orbits in the warped disc give rise to
asymmetric image morphologies, leading to the breakdown of many
simplifying assumptions in the untilted models. The observer azimuthal
viewing angle becomes an important new parameter, and the dependence
of observables on both the viewing geometry and the ion-electron
temperature ratio become non-trivial. The combination of these factors
greatly expands the possible parameter space for tilted disc models
relative to those assumed in all previous models of Sgr A*. 

There are numerous implications. The parameters of both the black hole
and the viewing geometry are poorly constrained (Figures
\ref{probdiststilti} and \ref{probtitemdot}) by mm-VLBI and spectral
observations. Previously reported parameter constraints 
\citep{broderick2009,brodericketal2011,moscibrodzka2009,dexter2009,dexteretal2010,shcherbakovetal2012}
only apply if the black hole spin or 
tilt values are extremely small. Upper limits from this work are $a <
0.3$ or $\beta < 15^\circ$, although qualitatively similar behavior
has been seen in a simulation with 
$\beta=10^\circ$ \citep{henisey2009}. The best-fitting
models (Table \ref{bfit}) to the mm data sort into two groups at low ($10^{-9}-10^{-8}
M_\odot \rm yr^{-1}$) and high ($\gtrsim 10^{-7} M_\odot \rm yr^{-1}$)
time-averaged mass accretion rates. 

The ``crescent'' black hole images
reported by \citet{dexteretal2010} are also found from these
best-fitting tilted images at $1.3$mm, especially when including the effects of interstellar
scattering. For this reason, the black hole shadow is still expected
to be accessible, although huge uncertainties in the viewing geometry
for these models prevent a robust prediction for baselines where it
should be seen. In two of the best-fitting models (see Figure
\ref{vis1dinterp}), the shadow is visible on baselines between
telescopes in Chile (ALMA/APEX) and Mexico (LMT), but a wide variety
of other geometries are nearly as likely. The mm variability properties reported previously
\citep{dexter2009,dexteretal2010} 
qualitatively hold for the tilted disc models, although their variability
amplitudes and characteristic timescales seem somewhat larger (Figure
\ref{lcurves}). 

Although images from tilted discs are crescents at $1.3$mm, at
$0.87$mm the images can become double-peaked due to emission from
near the standing shocks (915h model in Figure \ref{dint345}). The
structure of the tilted images is also highly time-variable, in stark
contrast to the static morphology in the untilted case (Figure
\ref{tiltuntiltcompare}). Tilted and untilted models can then be distinguished either
with higher frequency mm-VLBI or simply with many epochs of observations
using the current array. Polarized mm-VLBI observations may also help
to distinguish between models. \citet{shcherbakovetal2012} found that 
spectropolarimetric data contain comparable constraining power as
current mm-VLBI observations. 

Perhaps the most important difference is that the low accretion rate
best-fitting models (those from the 515h, 715h, and 915h simulations)
naturally reproduce IR spectral (Figure 
\ref{bfitspectra}) and variability (Figure \ref{lcurveshkl}) 
properties consistent with observations, while those at high mass
accretion rates (from the 315h simulation) underproduce the observed
flux. All models satisfy the observed upper limits on mid-infrared
emission \citep{telescoetal1996,coteraetal1999,schoedeletal2011}. All three low 
accretion rate models are in excellent agreement with the reported
mean NIR fluxes from Sgr A*
\citep{schoedeletal2011,doddsedenetal2011}. Flux distributions from
the 515h and 915h models are well described by a log-normal, and the
inferred parameters (Table \ref{irvar}) are in excellent agreement
with those found by \citet{doddsedenetal2011}. Both the mean and
time-dependence of the NIR spectral index varies considerably between
models. The 515h spectral index is constant in time with a value of
$\alpha_{\rm  NIR}=-0.5 \pm 0.1$, nearly identical to the observations of \citet{ghezetal2005} and
\citet{hornsteinetal2007}. The 715h and 915h light curves have much 
redder spectral indices, and the 715h spectral index increases with
flux. These values are in good agreement with 
\citet{gillessenetal2006}. Because of the short duration of our light
curves ($\simeq 25 \rm hr$), it is unclear whether the 515h, 715h, and 
915h light curves are best interpreted as different states of the same
theoretical model, or as a means to distinguish between the models.

In any event, this is the first model of Sgr A* that
self-consistently produces the observed time-variable mm to NIR
emission. The two are completely uncorrelated. The mm variations are due
to magnetic turbulence and are strongly correlated with the accretion
rate. The NIR variations are caused by particle density, field
strength, and temperature variations
in tenuous, hot gas heated by the non-axisymmetric standing
shocks, and can be significantly modulated by extreme gravitational
lensing. Short segments of these light curves 
can, however, still lead to the reported weak correlations (Figure
\ref{mmircrosscor}). We caution therefore that cross-correlation
coefficients of $\lesssim 0.6$ between wavebands in single epoch
observations do not necessarily imply a common physical origin.

\begin{figure}
\includegraphics[scale=.78]{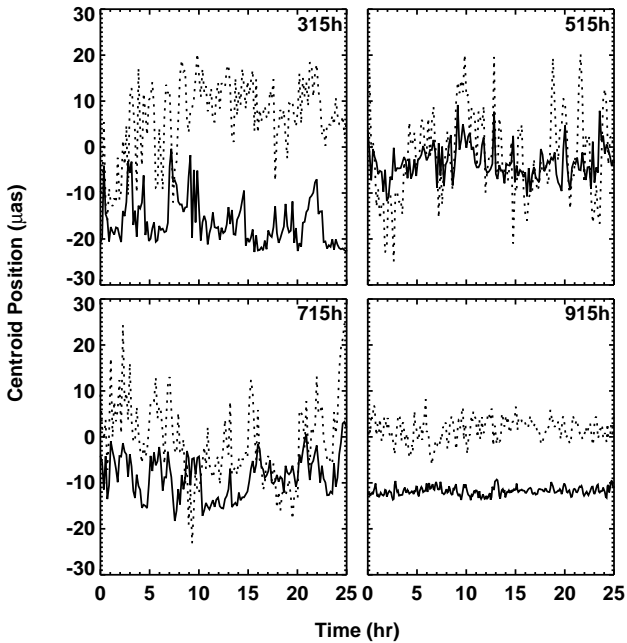}
\caption{\label{ircentroid}X (solid) and Y (dotted) centroid positions
for the K-band ($2.2 \mu$m) images. Periods of rapid centroid movement
of $\simeq 30-50 \mu$as occur every few hours in all models but 915h,
and should be detectable by the future VLT instrument GRAVITY.}
\end{figure}

\begin{figure}
\begin{tabular}{ll}
\includegraphics[scale=.8]{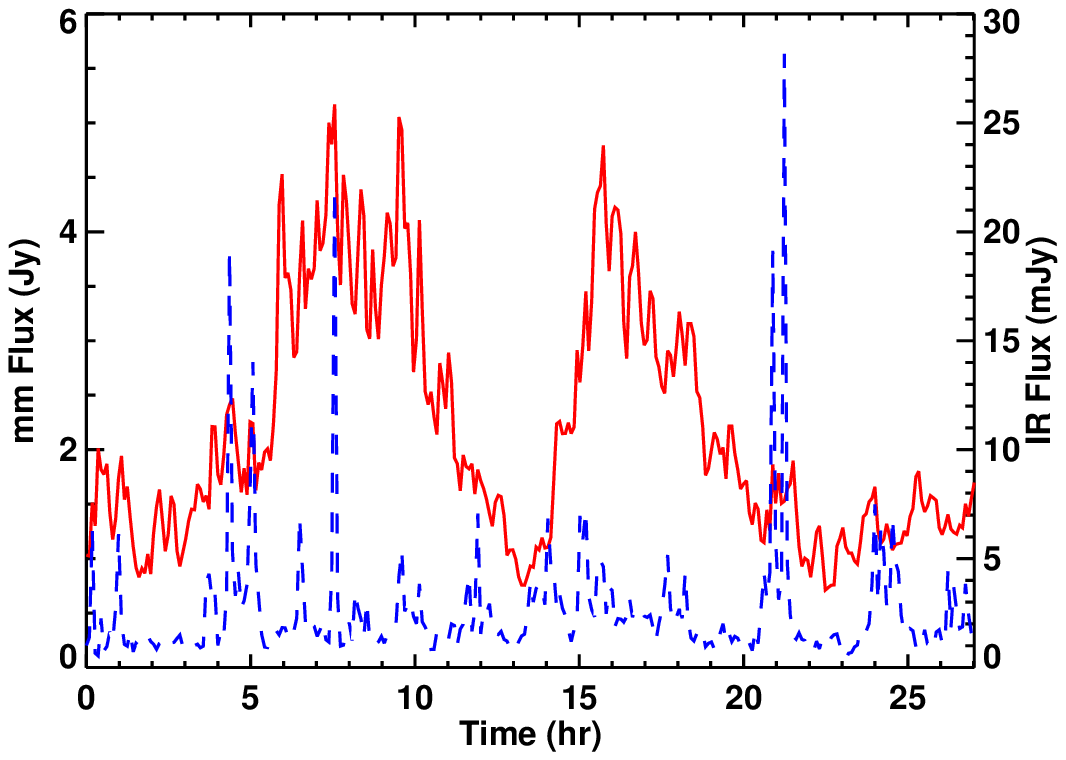}\\
\includegraphics[scale=.8]{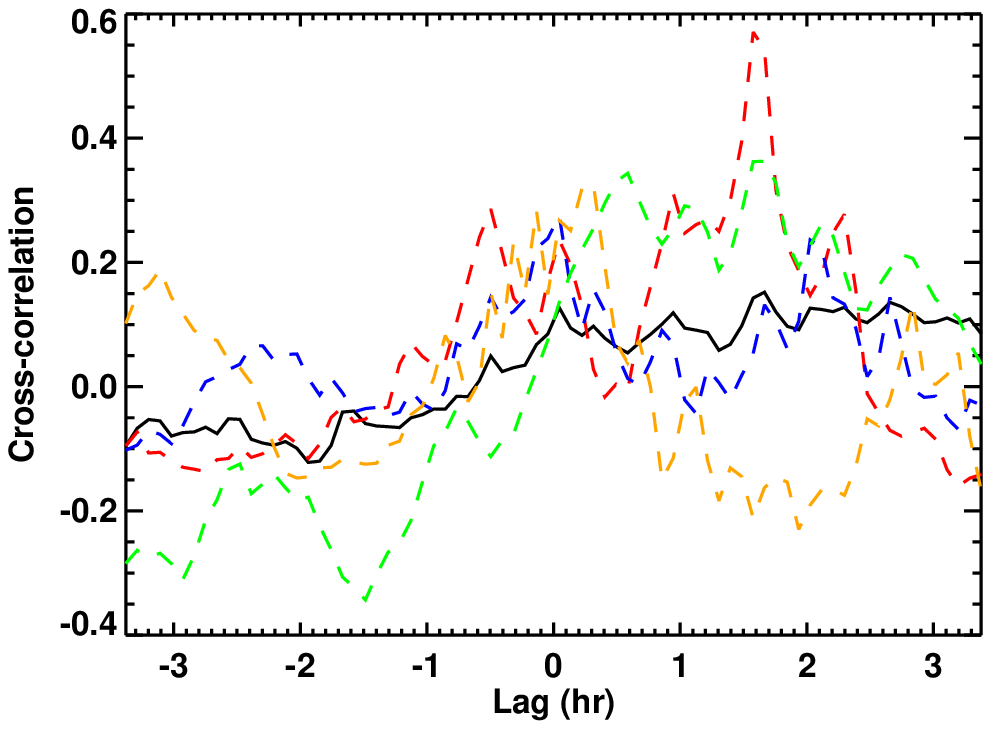}\\
\end{tabular}
\caption{\label{mmircrosscor}Top: Light curves from the 515h
  simulation at $0.4$mm (red) and $3.6\mu\rm m$ (blue). Bottom:
  Cross-correlation of the full light curves (solid curve) and four
  evenly spaced segments (dashed curves). For these short (several
  hour) durations, lags can be found at correlations of $\simeq 0.4-0.6$
  as reported in the literature \citep[e.g.,][]{yusefzadehetal2008},
  despite the fact that the full light curves are uncorrelated.}
\end{figure}

This NIR emission model is completely novel. It is not due to
orbiting or expanding ``hotspots''
\citep{broderickloeb2005,eckartetal2006,yusefzadehetal2008,eckartetal2012}. 
However, significant structural changes can still be seen during large
changes in NIR flux (Figure \ref{ircentroid}). It is similar in
spirit to models that invoke magnetic reconnection events
\citep{doddsedenetal2010} to 
accelerate electrons to the observed high energies. But magnetic
reconnection cannot heat particles in these simulations. Instead, the
heating is from the non-axisymmetric standing shocks in the tilted
accretion flow. This emission is also completely different from that described by
\citet{dolenceetal2012}, who detected quasi-periodic oscillations in
NIR/X-ray light curves of Sgr A* from an untilted GRMHD 
simulation. Their model underproduces the observed NIR flux by an order of
magnitude, similar to previous models from untilted simulations
\citep{dexteretal2010}. In addition, the NIR/X-ray variability in
their model is strongly correlated with that in the mm, in conflict
with observations. 

These radiative calculations assume that cooling is negligible for the
fluid (thermo)dynamics. However, the cooling timescale for the
NIR-emitting electrons is short:

\begin{equation}
  t_{\rm cool} \simeq 10 \left(\frac{B}{50 G}\right)^{-3/2}
  \left(\frac{\nu}{10^{14} \rm Hz}\right)^{-2} \rm min.
\end{equation}

\noindent At $r \approx 3 M$, where the emission is produced, this timescale is
still larger than the dynamical time ($t_{\rm dyn} \simeq 2$ min).
If this is where the particle heating occurs, ignoring the
effects of cooling on the electron temperature may be a safe
approximation. We have also assumed a thermal electron distribution
function, which is likely incorrect. Both of these 
limitations can be addressed in future calculations by 
following particle trajectories through the standing shocks and
self-consistently accounting for heating and cooling.

We have only included synchrotron radiation. Including inverse Compton
scattering via Monte Carlo calculations
\citep[e.g.,][]{dolenceetal2012} would allow the calculation of X-ray
light curves, and in particular the X-ray luminosity
during NIR flares. We can crudely estimate this using typical values
of fluid variables during rapid changes in NIR flux: $T_e =
2\times10^{12} \rm K$, $R = 1 M$ and $n=10^{5} \rm cm^{-3}$ (low,
H-band flux of 1 mJy) to $R= 3 M$ and $n=6 \times 10^{5} \rm
cm^{-3}$ (high, H-band flux of 10 mJy). For this temperature, the
Compton peak is at $\sim 1 \rm keV$. The ratios of Compton to
synchrotron bolometric luminosities 
estimated using the formulae in \citet{esinetal1996} are 
$L_{\rm C}/L_S \simeq 0.1$ (low) and $L_{\rm C}/L_S \simeq 2$
(high). At those H-band fluxes, the estimated X-ray luminosities are: 
$L_{\rm X} \sim 10^{33} \rm erg~\rm s^{-1}$ (low) and $L_{\rm X} \sim
10^{35} \rm erg~\rm s^{-1}$ (high). These numbers are in rough
agreement with observations of X-ray flares
\citep[e.g.,][]{baganoff2001,baganoff2003}. 

The effects of radiative cooling on
multi-wavelength spectra from axisymmetric (aligned) GRMHD simulations
have been studied by Dibi et al. (2012, submitted) and
\citet{drappeauetal2011}. They find that adding radiation in
post-processing, as done here, is valid for accretion rates $\dot{M}
\lesssim 10^{-7} M_\odot \rm yr^{-1}$. All of our best-fitting models
satisfy this criterion (Table \ref{bfit}) except for that from 315h,
where the accretion rate is slightly larger than 
this value. Radiative cooling may be important for this and similar
models, although it is unknown whether this accretion rate
limit applies to tilted discs.

We have only considered four simulations, all with a tilt angle of
$\beta=15^\circ$. The non-conservative GRMHD algorithm used in these
simulations leads to a scale height $H/r \simeq 0.1$, whereas the
accretion flow onto Sgr A* may have $H/r \simeq 1$. We are unable to
probe the dependence of our results on either tilt angle or scale
height, both of which may change the formation and strength of the
standing shocks. It is unclear whether the models considered here are 
representative of the range of possibilities from tilted black hole
accretion discs. 

All of these simulations failed to conserve total
energy. We previously found that magnetic energy lost to grid-scale
numerical reconnection leads to a radiatively efficient effective
luminosity, $\sim 0.1 \dot{M} c^2$ \citep{dexter2009}. Although
untilted models based on these simulations give very similar results
to those that recapture this energy as heat \citep{dexteretal2010}, it
is unclear whether that is the case for tilted discs. The lack of
energy conservation means that the only form of entropy generation is
through shock heating. Numerical dissipation of magnetic fields would
provide another heat source, and could potentially lessen the dramatic
effects seen here that are caused by the standing shocks. This
issue can be addressed by future energy-conserving tilted
simulations.

If standing shocks forming from eccentric orbits in a tilted, warped
disc provide the heating mechanism for the observed NIR emission, then
the inner disc must be tilted with respect to the black hole spin
axis. Then the Galactic centre black hole must have non-zero spin. Furthermore, recent
general relativistic, semi-analytic models of geometrically thick tilted
accretion discs find that the inner disc aligns with the black hole
for $a < 0$ \citep{zhuravlevivanov2011}. Therefore, if this proposed
model for the NIR emission is correct, the black hole in Sgr A* must
have positive spin ($a > 0$).

\section{Summary}
\label{sec:summary}

There is no reason to expect alignment between the black hole spin
axis and the angular momentum axis of accreting gas in radiatively
inefficient sources. We have calculated time-variable mm to NIR
emission from the only published numerical models of misaligned
(``tilted'') black hole accretion discs and compared them to
observations of Sgr A*. Our main results are: 

\begin{enumerate}

\item Tilted disc models provide an excellent description of existing
  mm-VLBI and spectral observations of Sgr A*.

\item Previously reported parameter constraints from fitting models to
  mm-VLBI, spectral, and/or polarization measurements are 
  only valid if the spin or tilt value is very small (upper limits
  of $a < 0.3$, $\beta < 15^\circ$). 

\item Predicted images for mm-VLBI are still crescents as in the
  aligned case, and the black hole shadow,
  direct evidence for the existence of an event horizon,  is
  still potentially visible on baselines between Mexico and Chile. 

\item Heating from non-axisymmetric standing shocks can naturally
  produce the observed NIR emission. This emission is
  uncorrelated with the mm variability, which also agrees with
  observations and is strongly correlated with the mass accretion
  rate. The X-ray flares in this picture arise from Compton
  upscattering of NIR seed photons.

\item We predict several observational signatures of disc tilt: a)
  structural changes observable with mm-VLBI from strong
  asymmetry in the emission region 
  on hour (dynamical) and/or day-year (Lense-Thirring precession)
  timescales; b) bimodal structures in 345 GHz
  mm-VLBI images from the presence of standing
  shocks in the accretion flow; and c) NIR
  centroid motions of $30-50\mu$as from extreme gravitational lensing
  during flares. 

\item If the NIR emission is caused by heating from standing shocks in
  a tilted accretion flow, then the spin parameter of Sgr A* must
  satisfy $a > 0$. 

\end{enumerate}

\section*{acknowledgements}
JD thanks Eric Agol, Omer Blaes, Josh Dolence, and Eliot Quataert for useful
discussions. This work was partially supported by NSF grants
AST-0807385 and PHY11-25915, NASA grant 05-ATP05-96, and NASA Earth \&
Space Science Fellowship NNX08AX59H (JD). 

\footnotesize{
\bibliographystyle{mn2e}


\begin{thebibliography}{78}
\expandafter\ifx\csname natexlab\endcsname\relax\def\natexlab#1{#1}\fi

\bibitem[{{Aitken} {et~al}\mbox{.}(2000){Aitken}, {Greaves}, {Chrysostomou},
  {Jenness}, {Holland}, {Hough}, {Pierce-Price}, \& {Richer}}]{aitken2000}
{Aitken} D.~K., {Greaves} J., {Chrysostomou} A., {Jenness} T., {Holland} W.,
  {Hough} J.~H., {Pierce-Price} D., {Richer} J., 2000, \apjl, 534, L173

\bibitem[{{An} {et~al}\mbox{.}(2005){An}, {Goss}, {Zhao}, {Hong}, {Roy}, {Rao},
  \& {Shen}}]{an2005}
{An} T., {Goss} W.~M., {Zhao} J., {Hong} X.~Y., {Roy} S., {Rao} A.~P., {Shen}
  Z., 2005, \apjl, 634, L49

\bibitem[{{Anninos}, {Fragile} \& {Salmonson}(2005){Anninos}, {Fragile}, \&
  {Salmonson}}]{anninos2005}
{Anninos} P., {Fragile} P.~C., {Salmonson} J.~D., 2005, \apj, 635, 723

\bibitem[{{Baganoff} {et~al}\mbox{.}(2001){Baganoff}, {Bautz}, {Brandt},
  {Chartas}, {Feigelson}, {Garmire}, {Maeda}, {Morris}, {Ricker}, {Townsley},
  \& {Walter}}]{baganoff2001}
{Baganoff} F.~K. {et~al.}, 2001, \nat, 413, 45

\bibitem[{{Baganoff} {et~al}\mbox{.}(2003){Baganoff}, {Maeda}, {Morris},
  {Bautz}, {Brandt}, {Cui}, {Doty}, {Feigelson}, {Garmire}, {Pravdo}, {Ricker},
  \& {Townsley}}]{baganoff2003}
---, 2003, \apj, 591, 891

\bibitem[{{Balbus} \& {Hawley}(1991)}]{mri}
{Balbus} S.~A., {Hawley} J.~F., 1991, \apj, 376, 214

\bibitem[{{Bardeen}(1973)}]{bardeen1973}
{Bardeen} J.~M., 1973, in Black holes (Les astres occlus), {DeWitt} B.~S.,
  {DeWitt} C., eds., New York: Gordon and Breach, p. 215

\bibitem[{{Bardeen} \& {Petterson}(1975)}]{bardeenpetterson1975}
{Bardeen} J.~M., {Petterson} J.~A., 1975, \apjl, 195, L65+

\bibitem[{{Bower} {et~al}\mbox{.}(2006){Bower}, {Goss}, {Falcke}, {Backer}, \&
  {Lithwick}}]{bower2006}
{Bower} G.~C., {Goss} W.~M., {Falcke} H., {Backer} D.~C., {Lithwick} Y., 2006,
  \apjl, 648, L127

\bibitem[{{Bower} {et~al}\mbox{.}(2003){Bower}, {Wright}, {Falcke}, \&
  {Backer}}]{bower03}
{Bower} G.~C., {Wright} M.~C.~H., {Falcke} H., {Backer} D.~C., 2003, \apj, 588,
  331

\bibitem[{{Broderick} {et~al}\mbox{.}(2009){Broderick}, {Fish}, {Doeleman}, \&
  {Loeb}}]{broderick2009}
{Broderick} A.~E., {Fish} V.~L., {Doeleman} S.~S., {Loeb} A., 2009, \apj, 697,
  45

\bibitem[{{Broderick} {et~al}\mbox{.}(2011){Broderick}, {Fish}, {Doeleman}, \&
  {Loeb}}]{brodericketal2011}
---, 2011, \apj, 735, 110

\bibitem[{{Broderick} \& {Loeb}(2005)}]{broderickloeb2005}
{Broderick} A.~E., {Loeb} A., 2005, \mnras, 363, 353

\bibitem[{{Broderick} \& {Loeb}(2006)}]{broderickloeb2006}
---, 2006, \apjl, 636, L109

\bibitem[{{Chakrabarti}(1985)}]{chakrabarti1985}
{Chakrabarti} S.~K., 1985, \apj, 288, 1

\bibitem[{{Chan} {et~al}\mbox{.}(2009){Chan}, {Liu}, {Fryer}, {Psaltis},
  {{\"O}zel}, {Rockefeller}, \& {Melia}}]{chan2009}
{Chan} C., {Liu} S., {Fryer} C.~L., {Psaltis} D., {{\"O}zel} F., {Rockefeller}
  G., {Melia} F., 2009, \apj, 701, 521

\bibitem[{{Cotera} {et~al}\mbox{.}(1999){Cotera}, {Simpson}, {Erickson},
  {Colgan}, {Burton}, \& {Allen}}]{coteraetal1999}
{Cotera} A.~S., {Simpson} J.~P., {Erickson} E.~F., {Colgan} S.~W.~J., {Burton}
  M.~G., {Allen} D.~A., 1999, \apj, 510, 747

\bibitem[{{Cuadra} {et~al}\mbox{.}(2006){Cuadra}, {Nayakshin}, {Springel}, \&
  {Di Matteo}}]{cuadraetal2006}
{Cuadra} J., {Nayakshin} S., {Springel} V., {Di Matteo} T., 2006, \mnras, 366,
  358

\bibitem[{{Dexter}(2011)}]{dexter2011}
{Dexter} J., 2011, PhD thesis, University of Washington

\bibitem[{{Dexter} \& {Agol}(2009)}]{dexteragol2009}
{Dexter} J., {Agol} E., 2009, \apj, 696, 1616

\bibitem[{{Dexter}, {Agol} \& {Fragile}(2009){Dexter}, {Agol}, \&
  {Fragile}}]{dexter2009}
{Dexter} J., {Agol} E., {Fragile} P.~C., 2009, \apjl, 703, L142

\bibitem[{{Dexter} {et~al}\mbox{.}(2010){Dexter}, {Agol}, {Fragile}, \&
  {McKinney}}]{dexteretal2010}
{Dexter} J., {Agol} E., {Fragile} P.~C., {McKinney} J.~C., 2010, \apj, 717,
  1092

\bibitem[{{Dexter} \& {Fragile}(2011)}]{dexterfragile2011}
{Dexter} J., {Fragile} P.~C., 2011, \apj, 730, 36

\bibitem[{{Do} {et~al}\mbox{.}(2009){Do}, {Ghez}, {Morris}, {Yelda}, {Meyer},
  {Lu}, {Hornstein}, \& {Matthews}}]{doetal2009}
{Do} T., {Ghez} A.~M., {Morris} M.~R., {Yelda} S., {Meyer} L., {Lu} J.~R.,
  {Hornstein} S.~D., {Matthews} K., 2009, \apj, 691, 1021

\bibitem[{{Dodds-Eden} {et~al}\mbox{.}(2011){Dodds-Eden}, {Gillessen}, {Fritz},
  {Eisenhauer}, {Trippe}, {Genzel}, {Ott}, {Bartko}, {Pfuhl}, {Bower},
  {Goldwurm}, {Porquet}, {Trap}, \& {Yusef-Zadeh}}]{doddsedenetal2011}
{Dodds-Eden} K. {et~al.}, 2011, \apj, 728, 37

\bibitem[{{Dodds-Eden} {et~al}\mbox{.}(2010){Dodds-Eden}, {Sharma}, {Quataert},
  {Genzel}, {Gillessen}, {Eisenhauer}, \& {Porquet}}]{doddsedenetal2010}
{Dodds-Eden} K., {Sharma} P., {Quataert} E., {Genzel} R., {Gillessen} S.,
  {Eisenhauer} F., {Porquet} D., 2010, \apj, 725, 450

\bibitem[{{Doeleman} {et~al}\mbox{.}(2008){Doeleman}, {Weintroub}, {Rogers},
  {Plambeck}, {Freund}, {Tilanus}, {Friberg}, {Ziurys}, {Moran}, {Corey},
  {Young}, {Smythe}, {Titus}, {Marrone}, {Cappallo}, {Bock}, {Bower},
  {Chamberlin}, {Davis}, {Krichbaum}, {Lamb}, {Maness}, {Niell}, {Roy},
  {Strittmatter}, {Werthimer}, {Whitney}, \& {Woody}}]{doeleman2008}
{Doeleman} S.~S. {et~al.}, 2008, \nat, 455, 78

\bibitem[{{Dolence} {et~al}\mbox{.}(2012){Dolence}, {Gammie}, {Shiokawa}, \&
  {Noble}}]{dolenceetal2012}
{Dolence} J.~C., {Gammie} C.~F., {Shiokawa} H., {Noble} S.~C., 2012, \apjl,
  746, L10

\bibitem[{{Drappeau} {et~al}\mbox{.}(2011){Drappeau}, {Dibi}, {Dexter},
  {Markoff}, \& {Fragile}}]{drappeauetal2011}
{Drappeau} S., {Dibi} S., {Dexter} J., {Markoff} S., {Fragile} P.~C., 2011, in
  SF2A-2011: Proceedings of the Annual meeting of the French Society of
  Astronomy and Astrophysics, {G.~Alecian, K.~Belkacem, R.~Samadi, \&
  D.~Valls-Gabaud}, ed., pp. 563--565

\bibitem[{{Eckart} {et~al}\mbox{.}(2004){Eckart}, {Baganoff}, {Morris},
  {Bautz}, {Brandt}, {Garmire}, {Genzel}, {Ott}, {Ricker}, {Straubmeier},
  {Viehmann}, {Sch{\"o}del}, {Bower}, \& {Goldston}}]{eckartetal2004}
{Eckart} A. {et~al.}, 2004, \aap, 427, 1

\bibitem[{{Eckart} {et~al}\mbox{.}(2006){Eckart}, {Baganoff}, {Sch{\"o}del},
  {Morris}, {Genzel}, {Bower}, {Marrone}, {Moran}, {Viehmann}, {Bautz},
  {Brandt}, {Garmire}, {Ott}, {Trippe}, {Ricker}, {Straubmeier}, {Roberts},
  {Yusef-Zadeh}, {Zhao}, \& {Rao}}]{eckartetal2006}
---, 2006, \aap, 450, 535

\bibitem[{{Eckart} {et~al}\mbox{.}(2012){Eckart}, {Garc{\'{\i}}a-Mar{\'{\i}}n},
  {Vogel}, {Teuben}, {Morris}, {Baganoff}, {Dexter}, {Sch{\"o}del}, {Witzel},
  {Valencia-S.}, {Karas}, {Kunneriath}, {Straubmeier}, {Moser}, {Sabha},
  {Buchholz}, {Zamaninasab}, {Mu{\v z}i{\'c}}, {Moultaka}, \&
  {Zensus}}]{eckartetal2012}
---, 2012, \aap, 537, A52

\bibitem[{{Esin} {et~al}\mbox{.}(1996){Esin}, {Narayan}, {Ostriker}, \&
  {Yi}}]{esinetal1996}
{Esin} A.~A., {Narayan} R., {Ostriker} E., {Yi} I., 1996, \apj, 465, 312

\bibitem[{{Falcke} {et~al}\mbox{.}(1998){Falcke}, {Goss}, {Matsuo}, {Teuben},
  {Zhao}, \& {Zylka}}]{falcke1998}
{Falcke} H., {Goss} W.~M., {Matsuo} H., {Teuben} P., {Zhao} J., {Zylka} R.,
  1998, \apj, 499, 731

\bibitem[{{Falcke} \& {Markoff}(2000)}]{falckemarkoff2000}
{Falcke} H., {Markoff} S., 2000, \aap, 362, 113

\bibitem[{{Falcke}, {Melia} \& {Agol}(2000){Falcke}, {Melia}, \&
  {Agol}}]{falcke}
{Falcke} H., {Melia} F., {Agol} E., 2000, \apjl, 528, L13

\bibitem[{{Fish} {et~al}\mbox{.}(2011){Fish}, {Doeleman}, {Beaudoin},
  {Blundell}, {Bolin}, {Bower}, {Chamberlin}, {Freund}, {Friberg}, {Gurwell},
  {Honma}, {Inoue}, {Krichbaum}, {Lamb}, {Marrone}, {Moran}, {Oyama},
  {Plambeck}, {Primiani}, {Rogers}, {Smythe}, {SooHoo}, {Strittmatter},
  {Tilanus}, {Titus}, {Weintroub}, {Wright}, {Woody}, {Young}, \&
  {Ziurys}}]{fishetal2011}
{Fish} V.~L. {et~al.}, 2011, \apjl, 727, L36+

\bibitem[{{Fragile}(2009)}]{fragiletilt2009}
{Fragile} P.~C., 2009, \apjl, 706, L246

\bibitem[{{Fragile} \& {Anninos}(2005)}]{fragile2005}
{Fragile} P.~C., {Anninos} P., 2005, \apj, 623, 347

\bibitem[{{Fragile} \& {Blaes}(2008)}]{fragile2008}
{Fragile} P.~C., {Blaes} O.~M., 2008, \apj, 687, 757

\bibitem[{{Fragile} {et~al}\mbox{.}(2007){Fragile}, {Blaes}, {Anninos}, \&
  {Salmonson}}]{fragile2007}
{Fragile} P.~C., {Blaes} O.~M., {Anninos} P., {Salmonson} J.~D., 2007, \apj,
  668, 417

\bibitem[{{Fragile} {et~al}\mbox{.}(2009){Fragile}, {Lindner}, {Anninos}, \&
  {Salmonson}}]{fragileetal2009}
{Fragile} P.~C., {Lindner} C.~C., {Anninos} P., {Salmonson} J.~D., 2009, \apj,
  691, 482

\bibitem[{{Genzel} {et~al}\mbox{.}(2003){Genzel}, {Sch{\"o}del}, {Ott},
  {Eckart}, {Alexander}, {Lacombe}, {Rouan}, \& {Aschenbach}}]{genzel2003}
{Genzel} R., {Sch{\"o}del} R., {Ott} T., {Eckart} A., {Alexander} T., {Lacombe}
  F., {Rouan} D., {Aschenbach} B., 2003, \nat, 425, 934

\bibitem[{{Ghez} {et~al}\mbox{.}(2005){Ghez}, {Hornstein}, {Lu}, {Bouchez}, {Le
  Mignant}, {van Dam}, {Wizinowich}, {Matthews}, {Morris}, {Becklin},
  {Campbell}, {Chin}, {Hartman}, {Johansson}, {Lafon}, {Stomski}, \&
  {Summers}}]{ghezetal2005}
{Ghez} A.~M. {et~al.}, 2005, \apj, 635, 1087

\bibitem[{{Ghez} {et~al}\mbox{.}(2004){Ghez}, {Wright}, {Matthews}, {Thompson},
  {Le Mignant}, {Tanner}, {Hornstein}, {Morris}, {Becklin}, \&
  {Soifer}}]{ghez2004}
---, 2004, \apjl, 601, L159

\bibitem[{{Gillessen} {et~al}\mbox{.}(2006){Gillessen}, {Eisenhauer},
  {Quataert}, {Genzel}, {Paumard}, {Trippe}, {Ott}, {Abuter}, {Eckart},
  {Lagage}, {Lehnert}, {Tacconi}, \& {Martins}}]{gillessenetal2006}
{Gillessen} S. {et~al.}, 2006, \apjl, 640, L163

\bibitem[{{Goldston}, {Quataert} \& {Igumenshchev}(2005){Goldston}, {Quataert},
  \& {Igumenshchev}}]{goldston2005}
{Goldston} J.~E., {Quataert} E., {Igumenshchev} I.~V., 2005, \apj, 621, 785

\bibitem[{{Hawley}, {Guan} \& {Krolik}(2011){Hawley}, {Guan}, \&
  {Krolik}}]{hawleyetal2011}
{Hawley} J.~F., {Guan} X., {Krolik} J.~H., 2011, \apj, 738, 84

\bibitem[{{Henisey} {et~al}\mbox{.}(2009){Henisey}, {Blaes}, {Fragile}, \&
  {Ferreira}}]{henisey2009}
{Henisey} K.~B., {Blaes} O.~M., {Fragile} P.~C., {Ferreira} B.~T., 2009, \apj,
  706, 705

\bibitem[{{Hornstein} {et~al}\mbox{.}(2007){Hornstein}, {Matthews}, {Ghez},
  {Lu}, {Morris}, {Becklin}, {Rafelski}, \& {Baganoff}}]{hornsteinetal2007}
{Hornstein} S.~D., {Matthews} K., {Ghez} A.~M., {Lu} J.~R., {Morris} M.,
  {Becklin} E.~E., {Rafelski} M., {Baganoff} F.~K., 2007, \apj, 667, 900

\bibitem[{{Huang} {et~al}\mbox{.}(2009){Huang}, {Liu}, {Shen}, {Yuan}, {Cai},
  {Li}, \& {Fryer}}]{huang2009}
{Huang} L., {Liu} S., {Shen} Z., {Yuan} Y., {Cai} M.~J., {Li} H., {Fryer}
  C.~L., 2009, \apj, 703, 557

\bibitem[{{Ivanov} \& {Illarionov}(1997)}]{ivanov1997}
{Ivanov} P.~B., {Illarionov} A.~F., 1997, \mnras, 285, 394

\bibitem[{{Kumar} \& {Pringle}(1985)}]{kumarpringle1985}
{Kumar} S., {Pringle} J.~E., 1985, \mnras, 213, 435

\bibitem[{{Leung}, {Gammie} \& {Noble}(2011){Leung}, {Gammie}, \&
  {Noble}}]{leungetal2011}
{Leung} P.~K., {Gammie} C.~F., {Noble} S.~C., 2011, \apj, 737, 21

\bibitem[{{Markoff} {et~al}\mbox{.}(2001){Markoff}, {Falcke}, {Yuan}, \&
  {Biermann}}]{markoffetal2001}
{Markoff} S., {Falcke} H., {Yuan} F., {Biermann} P.~L., 2001, \aap, 379, L13

\bibitem[{{Marrone}(2006)}]{marronephd}
{Marrone} D.~P., 2006, PhD thesis, AA(Harvard University)

\bibitem[{{Marrone} {et~al}\mbox{.}(2008){Marrone}, {Baganoff}, {Morris},
  {Moran}, {Ghez}, {Hornstein}, {Dowell}, {Mu{\~n}oz}, {Bautz}, {Ricker},
  {Brandt}, {Garmire}, {Lu}, {Matthews}, {Zhao}, {Rao}, \&
  {Bower}}]{marrone2008}
{Marrone} D.~P. {et~al.}, 2008, \apj, 682, 373

\bibitem[{{Marrone} {et~al}\mbox{.}(2007){Marrone}, {Moran}, {Zhao}, \&
  {Rao}}]{marrone07}
{Marrone} D.~P., {Moran} J.~M., {Zhao} J.-H., {Rao} R., 2007, \apjl, 654, L57

\bibitem[{{McKinney} \& {Blandford}(2009)}]{mckinneyblandford2009}
{McKinney} J.~C., {Blandford} R.~D., 2009, \mnras, 394, L126

\bibitem[{{Melia} \& {Falcke}(2001)}]{meliafalcke2001}
{Melia} F., {Falcke} H., 2001, \araa, 39, 309

\bibitem[{{Meyer} {et~al}\mbox{.}(2008){Meyer}, {Do}, {Ghez}, {Morris},
  {Witzel}, {Eckart}, {B{\'e}langer}, \& {Sch{\"o}del}}]{meyeretal2008}
{Meyer} L., {Do} T., {Ghez} A., {Morris} M.~R., {Witzel} G., {Eckart} A.,
  {B{\'e}langer} G., {Sch{\"o}del} R., 2008, \apjl, 688, L17

\bibitem[{{Mo{\'s}cibrodzka} {et~al}\mbox{.}(2009){Mo{\'s}cibrodzka}, {Gammie},
  {Dolence}, {Shiokawa}, \& {Leung}}]{moscibrodzka2009}
{Mo{\'s}cibrodzka} M., {Gammie} C.~F., {Dolence} J.~C., {Shiokawa} H., {Leung}
  P.~K., 2009, \apj, 706, 497

\bibitem[{{Noble} {et~al}\mbox{.}(2007){Noble}, {Leung}, {Gammie}, \&
  {Book}}]{noble2007}
{Noble} S.~C., {Leung} P.~K., {Gammie} C.~F., {Book} L.~G., 2007, Class. and
  Quant. Gravity, 24, 259

\bibitem[{{Ohsuga}, {Kato} \& {Mineshige}(2005){Ohsuga}, {Kato}, \&
  {Mineshige}}]{ohsuga2005}
{Ohsuga} K., {Kato} Y., {Mineshige} S., 2005, \apj, 627, 782

\bibitem[{{Pang} {et~al}\mbox{.}(2011){Pang}, {Pen}, {Matzner}, {Green}, \&
  {Liebend{\"o}rfer}}]{pangetal2011}
{Pang} B., {Pen} U.-L., {Matzner} C.~D., {Green} S.~R., {Liebend{\"o}rfer} M.,
  2011, \mnras, 415, 1228

\bibitem[{{Quataert}(2004)}]{quataert2004}
{Quataert} E., 2004, \apj, 613, 322

\bibitem[{{Riquelme} {et~al}\mbox{.}(2012){Riquelme}, {Quataert}, {Sharma}, \&
  {Spitkovsky}}]{riquelmeetal2012}
{Riquelme} M.~A., {Quataert} E., {Sharma} P., {Spitkovsky} A., 2012, ArXiv
  e-prints

\bibitem[{{Sch{\"o}del} {et~al}\mbox{.}(2011){Sch{\"o}del}, {Morris}, {Muzic},
  {Alberdi}, {Meyer}, {Eckart}, \& {Gezari}}]{schoedeletal2011}
{Sch{\"o}del} R., {Morris} M.~R., {Muzic} K., {Alberdi} A., {Meyer} L.,
  {Eckart} A., {Gezari} D.~Y., 2011, \aap, 532, A83

\bibitem[{{Sharma} {et~al}\mbox{.}(2007){Sharma}, {Quataert}, {Hammett}, \&
  {Stone}}]{sharma2007e}
{Sharma} P., {Quataert} E., {Hammett} G.~W., {Stone} J.~M., 2007, \apj, 667,
  714

\bibitem[{{Shcherbakov}, {Penna} \& {McKinney}(2010){Shcherbakov}, {Penna}, \&
  {McKinney}}]{shcherbakovetal2012}
{Shcherbakov} R.~V., {Penna} R.~F., {McKinney} J.~C., 2010, ArXiv e-prints

\bibitem[{{Shiokawa} {et~al}\mbox{.}(2012){Shiokawa}, {Dolence}, {Gammie}, \&
  {Noble}}]{shiokawaetal2012}
{Shiokawa} H., {Dolence} J.~C., {Gammie} C.~F., {Noble} S.~C., 2012, \apj, 744,
  187

\bibitem[{{Telesco}, {Davidson} \& {Werner}(1996){Telesco}, {Davidson}, \&
  {Werner}}]{telescoetal1996}
{Telesco} C.~M., {Davidson} J.~A., {Werner} M.~W., 1996, \apj, 456, 541

\bibitem[{{Yuan}, {Quataert} \& {Narayan}(2003){Yuan}, {Quataert}, \&
  {Narayan}}]{yuanquataert2003}
{Yuan} F., {Quataert} E., {Narayan} R., 2003, \apj, 598, 301

\bibitem[{{Yusef-Zadeh} {et~al}\mbox{.}(2006){Yusef-Zadeh}, {Bushouse},
  {Dowell}, {Wardle}, {Roberts}, {Heinke}, {Bower}, {Vila-Vilar{\'o}},
  {Shapiro}, {Goldwurm}, \& {B{\'e}langer}}]{yusefzadehetal2006}
{Yusef-Zadeh} F. {et~al.}, 2006, \apj, 644, 198

\bibitem[{{Yusef-Zadeh} {et~al}\mbox{.}(2009){Yusef-Zadeh}, {Bushouse},
  {Wardle}, {Heinke}, {Roberts}, {Dowell}, {Brunthaler}, {Reid}, {Martin},
  {Marrone}, {Porquet}, {Grosso}, {Dodds-Eden}, {Bower}, {Wiesemeyer},
  {Miyazaki}, {Pal}, {Gillessen}, {Goldwurm}, {Trap}, \&
  {Maness}}]{yusefzadeh2009}
---, 2009, \apj, 706, 348

\bibitem[{{Yusef-Zadeh} {et~al}\mbox{.}(2008){Yusef-Zadeh}, {Wardle}, {Heinke},
  {Dowell}, {Roberts}, {Baganoff}, \& {Cotton}}]{yusefzadehetal2008}
{Yusef-Zadeh} F., {Wardle} M., {Heinke} C., {Dowell} C.~D., {Roberts} D.,
  {Baganoff} F.~K., {Cotton} W., 2008, \apj, 682, 361

\bibitem[{{Zhao} {et~al}\mbox{.}(2003){Zhao}, {Young}, {Herrnstein}, {Ho},
  {Tsutsumi}, {Lo}, {Goss}, \& {Bower}}]{zhao2003}
{Zhao} J.-H., {Young} K.~H., {Herrnstein} R.~M., {Ho} P.~T.~P., {Tsutsumi} T.,
  {Lo} K.~Y., {Goss} W.~M., {Bower} G.~C., 2003, \apjl, 586, L29

\bibitem[{{Zhuravlev} \& {Ivanov}(2011)}]{zhuravlevivanov2011}
{Zhuravlev} V.~V., {Ivanov} P.~B., 2011, \mnras, 415, 2122

\end{thebibliography}
}
\label{lastpage}

\end{document}